\documentclass[10pt,preprint]{emulateapj}

\newcommand{\hst}{\textit{HST}}
\newcommand{\spitzer}{\textit{Spitzer}}

\newcommand{\acsb}{\hbox{$B_{435}$}}
\newcommand{\acsv}{\hbox{$V_{606}$}}
\newcommand{\acsi}{\hbox{$i_{775}$}}
\newcommand{\acsz}{\hbox{$z_{850}$}}
\newcommand{\nicj}{\hbox{$J_{110}$}}
\newcommand{\nich}{\hbox{$H_{160}$}}

\newcommand{\ks}{\hbox{$K_s$}}

\newcommand{\lsim}{\lesssim}
\newcommand{\gsim}{\gtrsim}
\newcommand{\lstar}{\hbox{$L^\ast$}}
\newcommand{\mcal}{\hbox{$\mathcal{M}$}}

\newcommand{\eg}{e.g.}

\newcommand{\msol}{\hbox{$\mathcal{M}_\odot$}}

\newcommand{\uJy}{\hbox{$\mu$Jy}}
\newcommand{\ujy}{\hbox{$\mu$Jy}}

\newcommand{\mone}{\hbox{$[3.6\mu\mathrm{m}]$}}
\newcommand{\mtwo}{\hbox{$[4.5\mu\mathrm{m}]$}}

\newcommand{\calL}{{\cal L}}
\newcommand \msun{\hbox{$\hbox{M}_{\odot}$}}

\begin{document}

\title{ON THE STELLAR POPULATIONS IN FAINT RED GALAXIES IN THE
\textit{HUBBLE} ULTRA DEEP FIELD}

\author{\sc Amelia M. Stutz, 
Casey Papovich\altaffilmark{2}, 
Daniel J. Eisenstein} 
\affil{Department of Astronomy and Steward Observatory, 
University of Arizona, 933 North Cherry Avenue, Tucson, 
Arizona 85721 
\\ astutz, papovich, eisenstein@as.arizona.edu}

\altaffiltext{1}{This work is based in part on observations taken with
  the NASA/ESA Hubble Space Telescope, which is operated by the
  Association of Universities for Research in Astronomy, Inc.\ (AURA)
  under NASA contract NAS5--26555; and on observations made with the
  \textit{Spitzer Space Telescope}, which is operated by the Jet
  Propulsion laboratory, California Institute of Technology, under
  NASA contract 1407. }
\altaffiltext{2}{\spitzer\ Fellow}

\setcounter{footnote}{2}

\begin{abstract}
We study the nature of faint, red-selected galaxies at $z \sim 2-3$
using the \textit{Hubble} Ultra Deep Field (HUDF) and \spitzer\
Infrared Array Camera (IRAC) photometry.  Given the magnitude limit of
the \hst\ data, we detect candidate galaxies to
$H_\mathrm{AB}<26$~mag, probing lower-luminosity (lower mass) galaxies
at these redshifts.  We identify 32 galaxies satisfying the
$(\nicj-\nich)_\mathrm{AB} > 1.0$~mag color selection, 16 of which
have unblended \mone\ and \mtwo\ photometry from \spitzer.  Using this
multiwavelength dataset we derive photometric redshifts, masses, and
stellar population parameters for these objects.  We find that the
selected objects span a diverse range of properties over a large range
of redshifts, $1 \lsim z \lsim 3.5$.  A substantial fraction (11/32)
of the $(\nicj - \nich)_\mathrm{AB} > 1$~mag population appear to be
lower-redshift ($z \lsim 2.5$), heavily obscured dusty galaxies or
edge-on spiral galaxies, while others (12/32) appear to be galaxies at
$2 \lsim z \lsim 3.5$ whose light at rest-frame optical wavelengths is
dominated by evolved stellar populations.  We argue that longer
wavelength data ($\gsim 1$~\micron, rest-frame) are essential for
interpreting the properties of the stellar populations in red-selected
galaxies at these redshifts.  Interestingly, by including \spitzer\
data many candidates for galaxies dominated by evolved stellar
populations are rejected, and for only a subset of the sample (6/16)
do the data favor this interpretation.  These objects have a surface
density of $\sim1$~arcmin$^{-2}$.  We place an upper limit on the
space density of candidate massive evolved galaxies with $2.5 < z <
3.5$ and $\nich(\mathrm{AB}) \leq 26$~mag of $n = 6.6^{+2.0}_{-3.0}
\times 10^{-4}$ Mpc$^{-3}$ with a corresponding upper limit on the
stellar mass density of $\rho^\ast = 5.6^{+4.4}_{-2.8}\times 10^{7}$
$M_\odot$ Mpc$^{-3}$.  The $z > 2.5$ objects that are dominated by
evolved stellar populations have a space density at most one-third
that of $z\sim 0$ red, early-type galaxies.  Therefore, at least
two-thirds of present-day early-type galaxies assemble or evolve into
their current configuration at redshifts below 2.5.  We find a dearth
of candidates for low-mass ($\lsim 2\times 10^{10}$~$\msun)$ galaxies
at $1.5 < z \lesssim 3$ that are dominated by passively evolving
stellar populations even though the data should be sensitive to them;
thus, at these redshifts, galaxies whose light is dominated by evolved
stellar populations are restricted to only those galaxies that have
assembled high stellar mass.

\end{abstract}

\keywords{cosmology: observations -- galaxies: evolution -- galaxies:
high-redshift -- galaxies: stellar content -- infrared: galaxies}

\section{Introduction}

In the hierarchical picture of galaxy evolution, galaxies assemble
within dark matter haloes that merge over time
\citep[\eg,][]{freedman01,sp03}.  This model has been highly
successful at reproducing galaxy clustering properties based on
well-understood gravitational physics \citep[\eg,][]{weinberg04}.
However, it is difficult to test how galaxies assemble their stellar
content within the context of these models because of uncertain
physics coupling the dark matter and baryonic matter in these haloes.

Observationally, the cosmic star-formation-rate (SFR) density rises by
roughly an order of magnitude from $z\sim 0$ to 1 and remains roughly
constant from $z\sim 1$-6 \citep[and references therein]{hopkins04}.
At the same time, roughly half of the stellar-mass density in galaxies
formed between $z\sim $3 and 1, corresponding to this peak in the SFR
density \citep[\eg,][]{dickinson03,rudnick03,glazebrook04}.  While
theoretical models broadly reproduce this evolution in the global
galaxy population \citep[\eg,][]{somerville01,hernquist03}, they have
difficulty in reproducing star-formation trends in individual
galaxies, particularly in accounting for the formation and evolution
of specific SFRs of the most massive objects
\citep{baugh03,somerville04,delucia06,croton06}.  In particular, it is
difficult for models to produce galaxies (at any redshift) that are
mostly devoid of current star-formation and evolve only passively
\citep[\eg,][]{croton06,dave06}.  Quantifying the number of passively
evolving galaxies --- and as a function of redshift and stellar mass
--- is helpful for constraining the processes that form stars within
galaxies.

Searches for high-redshift galaxies that are dominated by older,
passively evolving stellar populations are challenging.  The light
from these stars peaks at rest-frame optical and near-IR wavelengths
($\simeq 0.4-2$~\micron), which at $z\gsim 1$ are shifted into the
observed-frame near-IR.  Several surveys have used deep, near-IR
observations to search for galaxies up to $z\sim 1- 2$ whose light is
dominated by older stellar populations \citep[see \eg,][]{mccarthy04}.
Such objects should have very red $R-K$ colors, which span the
4000\AA/Balmer--break at these redshifts; several searches have
focused on so-called extremely red objects (EROs), with selected
objects typically satisfying $(R-K)_\mathrm{Vega} \gsim 5$~mag.  As a
class, EROs include both passively evolving early-type galaxies up to
$z\sim 1$ \citep[\eg,][]{cimatti02} and dust-reddened starbursts
\citep[\eg,][]{smail02}.  However, with only optical and near-IR
photometry, it was not possible to uniquely interpret the majority of
these objects \citep[\eg,][]{moustakas04,wilson07}.  Recent
observations at longer wavelengths from \spitzer\ indicate that more
than half of the ERO population have strong emission in the thermal IR
\citep{wilson04,yanl04b}, suggesting that such objects constitute a
significant portion of the red-selected galaxy population.

Recently, observers have used deep near-IR data to study red-selected
galaxies at higher redshifts.  \citet{franx03} used deep VLT/ISAAC
$JH\ks$ observations to identify a population of ``distant red
galaxies'' (DRGs) with $(J_s - \ks)_\mathrm{Vega} > 2.3$~mag, which
should select galaxies that have a strong Balmer/4000~\AA\ break
between the $J_s$ and \ks\ bands at $z\sim 2$-3.5 \citep[see
also][]{saracco01}.  Like EROs, the DRG color selection is also
sensitive to high redshift ($z\gsim 1$) galaxies dominated by
dust-enshrouded starbursts also have red observed $J-K$ colors
\citep[\eg,][]{smail02}.  Indeed, subsequent study of DRGs has shown
that the majority are massive galaxies that are actively forming stars
at $z\sim 1.5$-3
\citep{vd03,fs04,rubin04,knudsen05,reddy05,papovich06,webb06}, and
only a small subset appear to be completely devoid of star formation
and passively evolving \citep{labbe05,kriek06}.

However, nearly all surveys of red-selected objects have been
restricted to using near-IR images from ground-based telescopes, which
in practice limits these searches to $K_\mathrm{AB}\lsim 24$~mag
\citep[\eg,][]{labbe03}.  For passively evolving stellar populations
at $z\sim 3$, this magnitude limit acts as a limit in \textit{stellar
mass} of $\gsim 10^{11}$~\msol.  In contrast, the $z\sim 2-3.5$
UV-selected, star-forming galaxies (so-called Lyman-break galaxies,
LBGs) that dominate the SFR density \citep{reddy05} have typical
stellar masses $\sim 10^{10}$~\msol\ (for ``\lstar'' LBGs; Papovich et
al.\ 2001).  The inferred ages for LBGs are generally less than 1~Gyr
\citep{shapley01,shapley05} --- significantly less than the age of the
Universe at these epochs.  Thus, it is unknown if there exists a
population of ``faded'' LBGs, identifiable as red, passively evolving
galaxies with stellar masses $\sim 10^{10}$~\msol.  Deriving
constraints on the density of such objects would improve our
understanding of the evolution of these galaxies and on how galaxies
assemble their stellar content at these early epochs.  However,
detecting lower-mass, red galaxies at high redshift requires near-IR
surveys with higher sensitivity than what is practically available
from the ground.

Here, we use observations in the \textit{Hubble} Ultra Deep Field
\citep[HUDF; Beckwith et al.\ 2006 and][]{thompson05}.  This field has
the deepest optical \hst/ACS images to date, which combined with
extremely deep, near-IR \hst/NICMOS, and Spitzer (3.6, 4.5~\micron)
datasets form part of the Great Observatories Origins Deep Survey
(GOODS).  The depth of the images in this field provides a means to
explore high-redshift galaxies to a relatively low limiting
stellar-mass.  Here, we focus on a sample of galaxies selected with
red $\nicj - \nich$ colors in an effort to identify candidates for
passively evolving galaxies at $z\gsim 2$.  Similarly,
\citet{brammer07} use a $J-H$ color selection to identify red galaxies
at $z > 2$ using the deep FIRES near-IR survey imaging.  The
\hst/NICMOS data for the HUDF allow us to probe objects with lower
stellar masses than has been previously possible.  For example,
\citet{dickinson00} used deep NICMOS data in the \textit{Hubble} Deep
Field \citep[HDF;][]{williams96} to study an unusually red ($\nicj -
\nich \gsim 2$~mag) source, possibly a passively evolving $z\sim 3$
galaxy with $\mcal > 3\times 10^{10}$~\msol.  Similarly, Chen \&
Marzke (2004) used the HUDF \hst\ images to identify red galaxies
($\acsi-\nich > 2$) with photometric redshifts $z_\mathrm{ph} > 2.8$
(see also Yan et al.\ 2004).  In this work we use the HUDF data to
study the number density and mass density of objects at $2.5 < z <
3.5$ with colors consistent with older passively evolving stellar
populations.  To study the HUDF $\nicj - \nich > 1$ sources, when
possible we include GOODS IRAC 3.6-4.5~\micron\ and MIPS 24~\micron\
data, which allows us to better constrain the nature of the stellar
populations in these objects (as we discuss below).

Throughout this paper we assume the following cosmological parameters:
$\Omega_M =$ 0.3, $\Omega_{\Lambda} =$ 0.7 and H$_0$ = 70 km s$^{-1}$
Mpc$^{-1}$ \citep[consistent with the latest \textit{WMAP} results,
e.g.,][]{sp03}.  We use \acsb\acsv\acsi\acsz\nicj\nich\ to denote
magnitudes derived in the the \hst\ ACS and NICMOS bandpasses F435W,
F606W, F775W, F850LP, F110W and F160W respectively.  We denote
magnitudes derived from the IRAC channel 1 and channel 2 data as
$[3.6\micron]$ and $[4.5\micron]$.  For reference, the effective
wavelengths of these bandpasses are $0.43$, 0.60, 0.77, 0.91, 1.1,
1.6, 3.6 and 4.5~\micron. All magnitudes henceforth are on the AB
system, $m(\mathrm{AB}) = 23.9 - 2.5\log(f_\nu / 1\;\ujy)$.

\section{Red galaxies in the HUDF: photometry and selection}\label{section:data}

We combine \hst/ACS and NICMOS observations of the HUDF (Beckwith et
al.\ 2006\footnote{http://www.stsci.edu/hst/udf}; Thompson et al.\
2005) over a $2.4 \times 2.4$ arcmin$^2$ area with \spitzer\ IRAC
observations from GOODS (M.~Dickinson et al., in
preparation)\footnote{http://www.stsci.edu/science/goods/}.  The plate
scale of the drizzled images is $0.03$, $0.09$ and $0.6$ arcsec
pix$^{-1}$ for the ACS, NICMOS, and IRAC data respectively.  The $10
\sigma$ AB point source detection limits for the ACS and NICMOS images
are $29.2$, $30.0$, $29.7$, $28.7$, $27.0$ and $27.0$~mag for the
\acsb\acsv\acsi\acsz\nicj\nich\ images respectively.  For the IRAC
bands, the $5 \sigma$ AB point source detection limits are 26.3~mag
($0.11\mu$Jy) and 25.6~mag ($0.2\mu$Jy), for $[3.6\micron]$ and
$[4.5\micron]$ (M.~Dickinson, private communication, 2005).

\begin{figure}[t]
  \begin{center}
  \includegraphics[angle=0,width=0.49\textwidth]{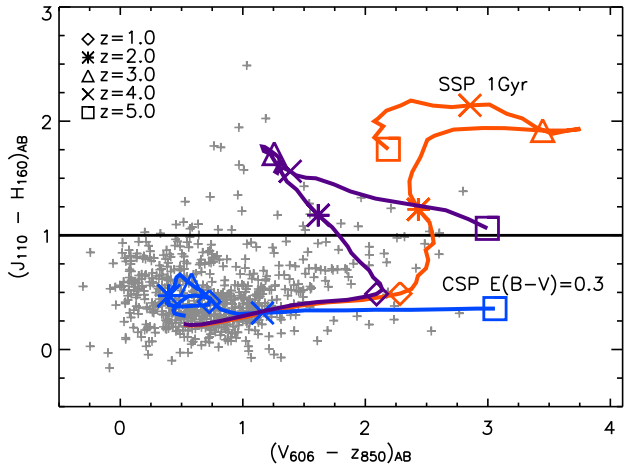}  
  \caption{ Color-color plot of all catalog objects with $H_{160} <
    26.0$.  The $J_{110} - H_{160} > 1.0$ color cut is marked.  The
    color evolution of three \citet{bc03} models is shown, with
    integer red-shift steps marked.  The curve labeled SSP (orange) is
    a $1.0$Gyr old simple stellar population formed in a burst with no
    subsequent star formation.  The CSP curve (blue) is a $100$Myr
    constant star-forming model attenuated using \citet{dc00} dust.
    The purple curve a is superposition of these two SSP and CSP
    models, with a ratio (SSP:CSP) of $99:1$ by mass.  Although the
    SSP and CSP models do not individually characterize the data well,
    the two-component model does.  The median error for the $J_{110} -
    H_{160}$ color is smaller than the symbol size; the median error
    for the $V_{606} - z_{850}$ color is similar to the symbol
    size.\label{fig1_1}}
  \end{center}
\end{figure}

We use ACS images reprocessed by \citet{thompson05} and matched to the
NICMOS plate-scale and orientation.  We then convolve the ACS images
by a kernel to match the point spread function (PSF) of the \nich-band
image.  Because the PSFs of the \nicj\ and \nich\ images are similar
we do not convolve the \nicj-band image.  We use standard IRAF
packages to perform the image convolution.  Our tests show that
point-source photometry on the convolved images recovers the same
fraction of the total flux to within $5\%$ for aperture radii greater
than $5$ pixels ($0.45$ arcsec).  Source detection was executed using
the SExtractor software \citep[v2.3.2,][]{bertin96} on a summed \nicj\
+ \nich\ image, and photometry was performed on the individual
PSF-matched images.  We use the same detection and photometry
parameters as in \citet{thompson05}.  We restrict our analysis to
objects with $\nich < 26$~mag measured in the SExtractor elliptical,
quasi-total (``Kron''; SExtractor MAG\_AUTO) apertures.  The NICMOS
data detect galaxies to 27th magnitude, and adopting a $\nich <
26$~mag limit ensures that we can derive robust colors for objects
with $\nicj - \nich = 1$.  We measure ACS and NICMOS colors for
objects in the PSF-matched images using isophotal (MAG\_ISO) apertures
derived from the \nich-image; we correct the isophotal magnitudes to
total magnitudes using the difference between the \nich-band MAG\_AUTO
and MAG\_ISO magnitudes.

The photometric errors from SExtractor are underestimated because they
do not take into account pixel-to-pixel correlations from, for
example, the drizzling and PSF-convolution processes.  We therefore
estimate photometric errors from the binned, drizzled, PSF-convolved
ACS images themselves.  We measure the sky noise as a function of
aperture size in blank regions of each image.  We then parameterize
the noise as a power law, $\sigma = A\times n_\mathrm{pix} ^\beta$,
where $\sigma$ is the noise measured in an aperture with a size of
$n_\mathrm{pix}$ pixels, and we fit for $\beta$ and $A$.  In order to
measure the noise in the convolved images we tabulate 1250 randomly
positioned aperture fluxes; we then bin these data and fit for the
width of the distribution with a Gaussian function.  We exclude bins
in the non-Gaussian positive tail of the distribution (these
correspond to apertures that receive some flux from bonafide sources
in the image).  This procedure is repeated for 11 aperture diameters
in each band: 0.54, 0.60, 0.80, 0.99, 1.15, 1.30, 1.40, 1.53, 1.65,
1.80 and 2$\farcs$25.  We obtain best-fit values of $\beta = 0.74$,
$0.92$, $0.96$, $0.93$ and $A = (5.6, 3.5, 2.0, 2.2)\times-10^{-4}$
ADU sec$^{-1}$, for \acsb\acsv\acsi\acsz\ respectively.  Using these
fitted values, we can then derive the error, $\sigma$ in each band for
any isophotal aperture size, $n_\mathrm{pix}$.

In figure~\ref{fig1_1} we plot the $J_{110} - H_{160}$ vs. $V_{606} -
z_{850}$ colors for all $H_{160} < 26.0$ catalog objects.  We also
show the color evolution of three \citet{bc03} models, a passively
evolving burst (SSP) model, a constant star-forming (CSP) model, and a
composite SSP+CSP model, at fixed age and as a function of redshift.
The composite model characterizes the $\nicj - \nich > 1$ selected
objects well.

\begin{figure}[t]
  \includegraphics[angle=0,width=0.49\textwidth]{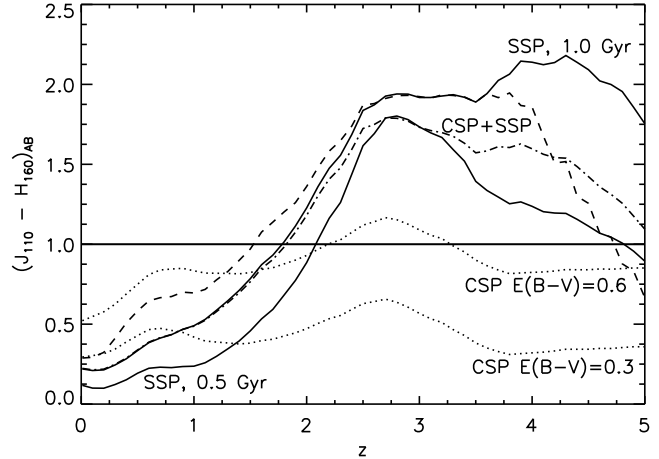}   
  \caption{ Color vs.\ redshift diagram of \citet{bc03} population
      synthesis models showing $J_{110} - H_{160}$ color evolution for
      different ages, star-formation histories and dust attenuations.
      The solid curves show a simple stellar population (SSP) model at
      two different ages ($1.0$ Gyr and $0.5$ Gyr).  The dotted curves
      show a $100$ Myr constant star-formation model (composite
      stellar population: CSP) with two different dust attenuations:
      $E(B-V) = 0.3$ (lower dotted curve) and $E(B-V) = 0.6$ (upper
      dotted curve).  The dot-dashed curve (labeled SSP+CSP) shows an
      example of a two-component model: an old population ($1.0$ Gyr
      SSP) and a young population ($100$ Myr constant star-formation
      with $E(B-V) = 0.3$) are combined with a ratio (SSP:CSP) of
      $99:1$ by mass.  The dashed curve shows a maximally old SSP
      formed at $z = 7$.  Note that our selection criterion, $J_{110}
      - H_{160} > 1.0$, generally excludes objects at redshifts less
      than $z \sim 1.5$.\label{fig1}}
\end{figure}

To target galaxies whose light is dominated by evolved stellar
populations, we select objects with $\nicj - \nich > 1.0$~mag and
$\nich \leq 26$~mag, which in principle selects objects with a
prominent $4000$~\AA/Balmer break at $z \gsim 1.5$.  In
figure~\ref{fig1} we plot several fiducial stellar population
synthesis models from \citet{bc03}, with solar metallicity and a
Salpeter IMF.  Our color criterion excludes model stellar populations
with recent star formation (and low-to-moderate dust attenuations),
and should be sensitive to passively evolving, older, simple stellar
populations (SSPs), or dust-enshrouded objects with younger stellar
populations.  We find 32 objects satisfying $\nicj - \nich > 1.0$~mag
and $\nich \leq 26.0$~mag in the NICMOS-selected catalog; their
properties are summarized in table~\ref{tab1}.

\begin{deluxetable*}{lcccccccccccl}
\tabletypesize{\scriptsize}
\tablecaption{Photometric properties of the NICMOS-UDF J$_{110}$ - H$_{160} > 1.0$ sample}
\tablewidth{0pt}
\tablehead{
\colhead{NIC ID} 
& \colhead{R.A.} 
& \colhead{Decl.} 
& \colhead{$z_{tot}$} 
& \colhead{$H_{tot}$} 
& \colhead{$B - z$} 
& \colhead{$V - z$} 
& \colhead{$I - z$} 
& \colhead{$z - H$} 
& \colhead{$J - H$} 
& \colhead{[3.6$\micron$]} 
& \colhead{[4.5$\micron$]}
& \colhead{Alternate ID} \\
 \colhead{ }
& \colhead{[J2000]}
& \colhead{[J2000]}
& \colhead{AB}
& \colhead{AB}
& \colhead{AB}
& \colhead{AB}
& \colhead{AB}
& \colhead{AB}
& \colhead{AB}
& \colhead{AB}
& \colhead{AB}
& \colhead{}
}
\startdata
30$^{*}$ & 03:32:39.72 & -27:46:11.3 & 23.60 & 21.73 & 1.55 & 1.28 & 0.63 & 1.87 & 1.18 & 19.96 & 19.77 & 1$^{c}$  \\
66 & 03:32:39.32 & -27:46:23.6 & 25.41 & 24.00 & 1.48 & 0.64 & 0.32 & 1.41 & 1.10 & \nodata & \nodata & \nodata  \\
86 & 03:32:37.13 & -27:46:25.9 & 24.99 & 23.23 & 5.41 & 2.60 & 1.03 & 1.76 & 1.02 & \nodata & \nodata & 2$^{c}$  \\
88$^{*}$ & 03:32:38.01 & -27:46:26.3 & 25.68 & 23.30 & 2.27 & 1.50 & 0.71 & 2.37 & 1.43 & 21.59 & 21.44 & 3$^{c}$  \\
99 & 03:32:36.17 & -27:46:27.7 & 24.58 & 23.07 & 1.47 & 0.80 & 0.45 & 1.51 & 1.05 & \nodata & \nodata & \nodata  \\
111$^{*}$ & 03:32:41.75 & -27:46:28.9 & 27.12 & 25.53 & 1.22 & 0.99 & 0.24 & 1.59 & 1.21 & 24.44 & 24.48 & \nodata  \\
120$^{*}$ & 03:32:41.01 & -27:46:31.5 & 26.57 & 24.50 & 1.53 & 0.96 & 0.42 & 2.07 & 1.35 & 22.42 & 22.21 & \nodata  \\
215 & 03:32:34.77 & -27:46:46.1 & 26.41 & 25.08 & 1.61 & 0.91 & 0.23 & 1.33 & 1.05 & \nodata & \nodata & \nodata  \\
219$^{*}$ & 03:32:43.32 & -27:46:46.9 & 26.16 & 23.76 & 1.79 & 0.76 & 0.24 & 2.39 & 1.78 & 22.21 & 22.11 & 9151$^{a}$  \\
223$^{*}$ & 03:32:35.06 & -27:46:47.6 & 25.79 & 23.36 & 1.97 & 0.97 & 0.43 & 2.43 & 1.62 & 21.61 & 21.48 & 9024$^{a}$, 4$^{c}$, 8$^{d}$  \\
234 & 03:32:42.96 & -27:46:49.9 & 23.43 & 21.81 & 4.29 & 2.52 & 0.90 & 1.62 & 1.03 & \nodata & \nodata & \nodata  \\
245 & 03:32:39.88 & -27:46:50.0 & 26.67 & 25.71 & 0.07 & 0.20 & 0.04 & 0.97 & 1.03 & \nodata & \nodata & \nodata  \\
269$^{*}$ & 03:32:38.73 & -27:46:54.1 & 25.74 & 24.63 & 1.11 & 0.45 & 0.18 & 1.10 & 1.09 & 23.28 & 23.55 & \nodata  \\
281 & 03:32:41.67 & -27:46:55.4 & 26.03 & 23.72 & 2.05 & 1.54 & 0.97 & 2.31 & 1.28 & \nodata & \nodata & \nodata  \\
319$^{*}$ & 03:32:34.42 & -27:46:59.7 & 24.28 & 22.31 & 3.03 & 2.21 & 0.93 & 1.97 & 1.24 & 20.40 & 20.25 & 5$^{c}$  \\
478 & 03:32:42.52 & -27:47:14.3 & 26.14 & 24.75 & 1.77 & 0.52 & 0.04 & 1.39 & 1.12 & \nodata & \nodata & \nodata  \\
535 & 03:32:34.62 & -27:47:20.9 & 25.00 & 23.38 & 2.38 & 1.94 & 0.84 & 1.61 & 1.08 & \nodata & \nodata & \nodata  \\
625$^{*}$ & 03:32:43.46 & -27:47:27.4 & 26.63 & 24.34 & 2.63 & 0.82 & 0.18 & 2.29 & 1.77 & 23.18 & 23.26 & 6140$^{a}$  \\
683 & 03:32:36.96 & -27:47:27.2 & 25.22 & 23.23 & 2.33 & 1.49 & 0.62 & 1.99 & 1.08 & \nodata & \nodata & \nodata  \\
706$^{*}$ & 03:32:42.73 & -27:47:33.9 & 24.37 & 22.74 & 1.90 & 1.51 & 0.75 & 1.63 & 1.05 & 21.28 & 21.27 & 5256$^{a}$  \\
902 & 03:32:38.10 & -27:47:49.8 & 24.47 & 22.17 & 3.82 & 1.80 & 0.93 & 2.30 & 1.10 & \nodata & \nodata & 3650$^{b}$  \\
921$^{*}$ & 03:32:39.11 & -27:47:51.6 & 25.50 & 23.11 & 3.36 & 1.55 & 0.86 & 2.39 & 1.19 & 21.92 & 21.96 & 3574$^{b}$, 13$^{d}$  \\
935 & 03:32:33.26 & -27:47:52.4 & 27.23 & 24.77 & 2.05 & 1.97 & 0.98 & 2.46 & 1.33 & \nodata & \nodata & 6$^{d}$  \\
989$^{*}$ & 03:32:34.93 & -27:47:56.0 & 26.21 & 24.76 & 1.70 & 0.57 & 0.35 & 1.44 & 1.06 & 23.37 & 23.44 & \nodata  \\
1078$^{*}$ & 03:32:42.87 & -27:48:09.5 & 28.06 & 24.54 & 2.02 & 1.03 & 0.62 & 3.51 & 2.49 & 22.03 & 21.78 & 12182$^{a}$, 1$^{d}$  \\
1082 & 03:32:38.70 & -27:48:10.3 & 27.35 & 25.84 & 1.75 & 0.73 & 0.09 & 1.51 & 1.34 & \nodata & \nodata & \nodata  \\
1172$^{*}$ & 03:32:41.74 & -27:48:24.9 & 27.75 & 24.56 & 2.02 & 2.80 & 0.83 & 3.18 & 1.39 & 22.48 & 22.32 & 4$^{d}$  \\
1184 & 03:32:38.75 & -27:48:27.1 & 26.43 & 24.57 & 1.84 & 0.53 & 0.17 & 1.86 & 1.30 & \nodata & \nodata & 15$^{d}$  \\
1211$^{*}$ & 03:32:39.16 & -27:48:32.4 & 25.78 & 23.25 & 2.30 & 1.19 & 0.33 & 2.52 & 2.02 & 22.08 & 22.06 & 1927$^{a}$, 1446$^{b}$, 6$^{c}$, 9$^{d}$  \\
1237$^{*}$ & 03:32:38.73 & -27:48:39.9 & 27.07 & 24.48 & 2.37 & 0.97 & 0.38 & 2.59 & 1.95 & 22.03 & 21.82 & 12183$^{a}$, 2$^{d}$  \\
1245 & 03:32:39.58 & -27:48:42.2 & 26.21 & 25.25 & 1.49 & 0.59 & 0.15 & 0.95 & 1.07 & \nodata & \nodata & \nodata  \\
1267 & 03:32:39.66 & -27:48:50.6 & 24.36 & 22.38 & 2.96 & 1.01 & 0.22 & 1.98 & 1.57 & \nodata & \nodata & 1223$^{a}$  \\

\enddata
\label{tab1}
\tablecomments{Objects with unblended IRAC photometry are marked with
  $^*$.  References for the alternate ID's listed are (a)
  \citet{chen04}, (b) \citet{daddi05}, (c) \citet{toft05} and (d) \citet{yan04a}.}
\end{deluxetable*}

\begin{figure}[t]
  \includegraphics[angle=0,width=0.47\textwidth]{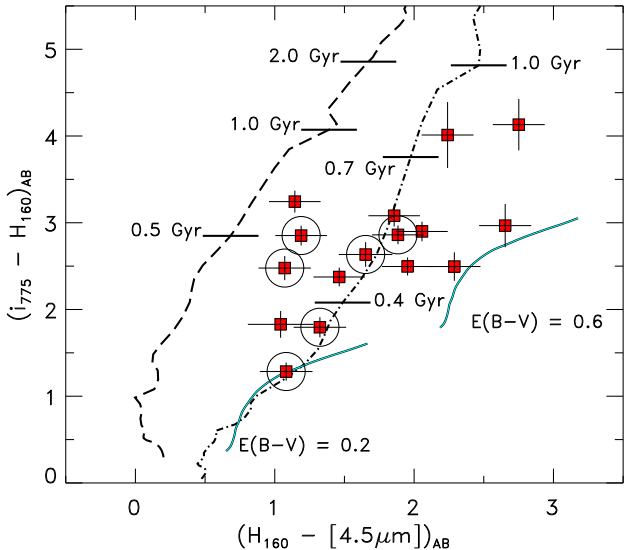}   
  \caption{ $\nich - \mtwo$ versus $\acsi - \nich$ colors for the 16
      galaxies in the $\nicj - \nich > 1$~mag sample with robust IRAC
      photometry.  The lines show colors of various model stellar
      populations formed at $z = 7$.  The dashed line shows the
      expected colors of a passively evolving stellar population
      observed at $z = 2.7$, formed in a burst as a function of age,
      $t$; the dot-dashed line shows the same model observed at $z =
      3.5$.  Tick marks denote the model ages along the dashed and
      dot-bashed curves as labeled.  The solid lines show colors of a
      stellar population with constant star formation formed at $z =
      7$ as a function of age and extinction with color excess $E(B-V)
      = 0.2$ and 0.6, as labeled.  The encircled data points indicate
      the 6 candidates for distant galaxies dominated by old stellar
      populations (see \S~4.1).  These models roughly bound the galaxy
      colors, and indicate that including the IRAC data we can
      understand the properties of the stellar populations in these
      galaxies.  Note that none of the objects have colors consistent
      with passively evolving stellar populations older than 1 Gyr.}
  \label{cfig1}
\end{figure}

We match $\nicj-\nich > 1$~mag objects against the \spitzer/IRAC
catalog from the Great Observatories Origins Deep Survey (GOODS;
M.~Dickinson et al. in preparation).  Source detection for this
catalog was performed on a summed $[3.6\micron$]+$[4.5\micron]$ image,
and photometry was measured in circular, 4.0\arcsec-diameter
apertures, which we correct to total magnitudes by adding $-0.36$ and
$-0.31$ respectively to the $[3.6\micron$] and $[4.5\micron]$
magnitudes (appropriate for GOODS, M. Dickinson et al., in prep, and
slightly larger than those in the IRAC data
handbook\footnote{http://ssc.spitzer.caltech.edu/irac/dh/}).  Because
the image PSF size is substantially different between the NICMOS and
IRAC images (FWHM $0\farcs3$ and $1\farcs6$, respectively), we do not
attempt to PSF match these images.

We matched these objects to sources in the \spitzer/IRAC catalog first
using a 4\arcsec-radius criterion.  We then visually inspected each
source and accept only those objects whose positions coincide with the
\hst\ source (they are well-centered) and who are isolated (unblended)
from other \hst\ sources within that radius.  This is a conservative
approach, and has the advantage of limiting our analysis to only those
\spitzer\ counterparts that are free from source confusion.  Of the 32
objects in the parent sample, 16 have unblended IRAC $[3.6\micron]$
and $[4.5\micron]$ counterparts.  In figures~\ref{fig2}-\ref{fig5} we
show the ACS, NICMOS, and IRAC images and spectral energy
distributions (SEDs) of these objects.

Including the IRAC data extends our wavelength coverage, allowing us
to discriminate broadly between galaxies dominated by dust-obscured
starbursts and galaxies with a substantial amount of stellar mass in
later-type stars (see, e.g., Labb\'e et al.\ 2005).  In
figure~\ref{cfig1}, we show a $\nich - \mtwo$ versus $\acsi-\nich$
plot to study the stellar populations and dust extinction properties
for the 16 galaxies with IRAC photometry.  We also plot fiducial
colors of a passively evolving stellar population and dust-enshrouded
starbursts (with $E(B-V) = 0.2$ and 0.6, using the extinction law from
Calzetti et al.\ 2000) as a function of stellar population age at
$z=2.7$.  These simple models bound the range of colors observed in
the $\nicj - \nich > 1$~mag objects, comparable to findings for red,
$J-\ks$-selected galaxies in other fields (Labb\'e et al.\ 2005;
Papovich et al.\ 2006).  Blue $\nich - \mtwo$ colors ($\lsim 2$~mag)
at this redshift (rest-frame $0.4-1.2$~micron) result from substantial
older stellar populations, which produce a significant
Balmer/4000~\AA\ break, causing redder $\acsi - \nich$ colors (and
also the red $\nicj - \nich$ color that satisfy the selection
criteria).  Objects at these redshifts with redder $\nich - \mtwo$
colors ($\gsim 2$~mag) require dust-enshrouded, ongoing star
formation.  The galaxies with $\nicj - \nich > 1$ and IRAC photometry
all have a similar range of $\acsi - \nich$ colors, making it
difficult to differentiate between the two types of stellar
populations.  However, these galaxies span a large range in $\nich -
\mtwo$ colors, from 1-4~mag.  Many of the objects appear to have
colors consistent with dust-enshrouded, young stellar populations,
while a few sources have colors consistent with a substantial amount
of their stellar mass in older stars.  Based on figure~3, there are
few galaxies whose colors are consistent with a pure passively
evolving, older stellar population.  We discuss this further in \S~4.

\begin{figure}[t]
    \includegraphics[angle=0,width=0.49\textwidth]{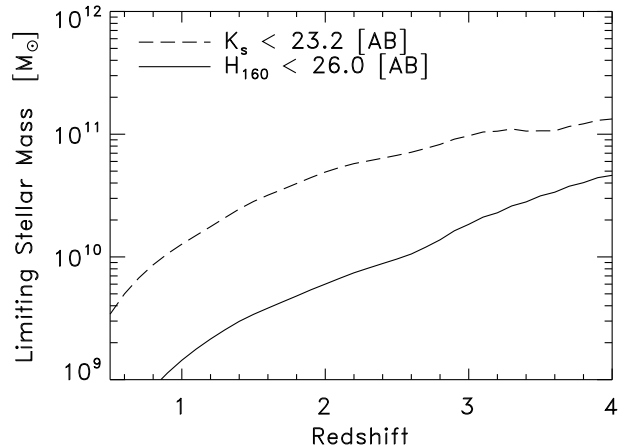}   
  \caption{Limiting stellar mass for galaxies with a stellar
    population formed in a burst at $z = 7$ with observed magnitude
    $\nich = 26.0$ or $\ks = 23.2$.  The HUDF data are sensitive to
    passively evolving stellar populations with masses above these
    limits. \label{fig:masslimit}}
  \label{cfig2}
\end{figure}

We also include other multiwavelength data to interpret the nature of
the $\nicj - \nich > 1$~mag sample.  In particular, we use
\spitzer/MIPS 24~\micron\ data for this field from the catalog of
Papovich et al.\ (2004).  These data allow us to test for the emission
from dust associated with either star formation or obscured AGN in
these galaxies.  We also include \textit{Chandra} X--ray data from the
catalog of Alexander et al.\ (2003).  At the redshifts of interest for
our sample, the X--ray data are sensitive accretion of material onto
supermassive blackholes, and thus allow us to test for the presence of
AGN in our sample.

\begin{deluxetable*}{lccccccc}
\tabletypesize{\scriptsize}
\tablecaption{Best-fit parameters with HST data only}
\tablewidth{0pt}
\tablehead{
\colhead{NIC ID} 
& \colhead{z$_{phot}$} 
& \colhead{E(B-V)$^{a}$} 
& \colhead{log M$_{o}$/M$_\sun$$^{b}$} 
& \colhead{log M$_{y}$/M$_\sun$$^{c}$} 
& \colhead{f$_{old}$$^{d}$} 
& \colhead{$\chi_{\nu,best}^{2}$$^{e}$}
& \colhead{$\Delta \chi^{2}$$^{f}$}
}
\startdata
30$^{*}$ & 1.4$^{+0.2}_{-0.1}$ & 0.2$\pm$0.1 &  11.2$\pm$0.1 & \phn9.0$^{+0.5}_{-0.3}$ & 0.993$^{+0.004}_{-0.010}$ & $\phn$5.2 & $\phn$0.0  \\
66 & 2.6$\pm$0.2 & 0.3$^{+0.1}_{-0.0}$ &  10.6$^{+0.1}_{-0.4}$ & \phn9.6$^{+0.2}_{-0.1}$ & 0.915$^{+0.021}_{-0.207}$ & $\phn$2.5 & $\phn$0.3  \\
86 & 1.1$^{+3.3}_{-0.1}$ & 1.1$^{+0.3}_{-0.8}$ &  10.3$^{+1.5}_{-0.2}$ & \phn9.8$^{+0.3}_{-0.5}$ & 0.763$^{+0.224}_{-0.136}$ & $\phn$0.0 & $\phn$0.8  \\
88$^{*}$ & 2.2$^{+0.1}_{-0.2}$ & 0.6$\pm$0.1 &  10.8$^{+0.1}_{-0.2}$ &  10.0$\pm$0.4 & 0.867$^{+0.083}_{-0.231}$ & $\phn$0.7 & $\phn$0.0  \\
99 & 2.1$\pm$0.1 & 0.4$\pm$0.0 &  10.7$\pm$0.1 & \phn10.$\pm$0.1 & 0.857$^{+0.018}_{-0.023}$ & $\phn$3.0 & $\phn$0.0  \\
111$^{*}$ & 0.7$^{+0.1}_{-0.0}$ & 0.0$^{+0.1}_{-0.0}$ & \phn8.8$\pm$0.0 & \phn6.5$^{+0.3}_{-0.0}$ & 0.996$^{+0.000}_{-0.005}$ & 17.9 & $\phn$0.0  \\
120$^{*}$ & 2.4$^{+0.3}_{-0.2}$ & 0.4$\pm$0.1 &  10.5$^{+0.2}_{-0.1}$ & \phn9.2$^{+0.3}_{-0.4}$ & 0.948$^{+0.032}_{-0.085}$ & $\phn$0.4 & $\phn$0.0  \\
215 & 0.6$\pm$0.0 & 0.5$\pm$0.0 & \phn6.6$^{+0.3}_{-0.7}$ & \phn7.9$\pm$0.0 & 0.048$^{+0.051}_{-0.034}$ & 12.6 & $\phn$0.0  \\
219$^{*}$ & 2.8$\pm$0.1 & 0.3$\pm$0.0 &  11.0$\pm$0.1 & \phn9.3$^{+0.0}_{-0.1}$ & 0.981$^{+0.004}_{-0.002}$ & $\phn$3.5 & $\phn$0.0  \\
223$^{*}$ & 2.9$^{+0.3}_{-0.4}$ & 0.3$\pm$0.1 &  11.3$^{+0.2}_{-0.3}$ & \phn9.4$^{+0.2}_{-0.3}$ & 0.986$^{+0.009}_{-0.027}$ & $\phn$1.3 & $\phn$1.1  \\
234 & 0.9$^{+0.1}_{-0.0}$ & 1.1$^{+0.2}_{-0.1}$ &  10.6$\pm$0.1 &  10.1$\pm$0.2 & 0.783$^{+0.094}_{-0.076}$ & $\phn$1.8 & $\phn$0.0  \\
245 & 2.4$\pm$0.1 & 0.0$\pm$0.0 & \phn10.$\pm$0.0 & \phn8.1$\pm$0.0 & 0.989$\pm$0.001 & 14.7 & $\phn$0.0  \\
269$^{*}$ & 2.7$^{+0.3}_{-0.1}$ & 0.2$^{+0.0}_{-0.1}$ &  10.4$^{+0.3}_{-0.1}$ & \phn9.2$^{+0.0}_{-0.2}$ & 0.941$^{+0.043}_{-0.020}$ & $\phn$2.3 & $\phn$0.0  \\
281 & 1.9$^{+0.2}_{-0.1}$ & 0.3$^{+0.2}_{-0.1}$ &  10.7$\pm$0.1 & \phn8.7$^{+0.6}_{-0.5}$ & 0.992$^{+0.005}_{-0.029}$ & $\phn$0.6 & $\phn$0.0  \\
319$^{*}$ & 1.1$\pm$0.1 & 0.9$^{+0.1}_{-0.0}$ &  10.1$^{+0.3}_{-0.5}$ &  10.2$\pm$0.2 & 0.421$^{+0.218}_{-0.200}$ & $\phn$1.9 & $\phn$0.0  \\
478 & 3.1$\pm$0.2 & 0.2$^{+0.1}_{-0.0}$ &  10.7$^{+0.1}_{-0.3}$ & \phn9.2$^{+0.2}_{-0.1}$ & 0.968$^{+0.007}_{-0.086}$ & $\phn$3.1 & $\phn$2.5  \\
535 & 1.1$\pm$0.1 & 0.7$^{+0.1}_{-0.5}$ &  10.0$^{+0.3}_{-0.5}$ & \phn9.3$^{+0.3}_{-1.5}$ & 0.788$^{+0.208}_{-0.330}$ & $\phn$4.4 & $\phn$0.0  \\
625$^{*}$ & 3.3$\pm$0.4 & 0.2$\pm$0.1 &  11.1$^{+0.2}_{-0.3}$ & \phn8.9$\pm$0.4 & 0.993$^{+0.005}_{-0.017}$ & $\phn$3.9 & $\phn$0.0  \\
683 & 1.6$^{+0.2}_{-0.1}$ & 0.6$\pm$0.1 &  10.5$^{+0.2}_{-0.3}$ & \phn9.9$^{+0.3}_{-0.5}$ & 0.800$^{+0.139}_{-0.291}$ & $\phn$0.6 & $\phn$0.0  \\
706$^{*}$ & 1.2$^{+0.1}_{-0.0}$ & 0.3$^{+0.1}_{-0.2}$ &  10.6$\pm$0.1 & \phn8.7$^{+0.4}_{-0.6}$ & 0.988$^{+0.009}_{-0.022}$ & $\phn$3.5 & $\phn$0.7  \\
902 & 1.9$\pm$0.1 & 0.8$\pm$0.1 &  11.1$^{+0.1}_{-0.3}$ &  10.8$^{+0.2}_{-0.4}$ & 0.682$^{+0.183}_{-0.295}$ & $\phn$2.8 & $\phn$0.0  \\
921$^{*}$ & 2.0$^{+0.1}_{-0.2}$ & 0.7$\pm$0.1 &  10.8$^{+0.1}_{-0.3}$ &  10.2$^{+0.4}_{-0.5}$ & 0.811$^{+0.124}_{-0.306}$ & $\phn$1.3 & $\phn$0.0  \\
935 & 1.9$\pm$0.2 & 0.5$\pm$0.4 &  10.3$^{+0.1}_{-0.3}$ & \phn8.6$\pm$1.2 & 0.978$^{+0.021}_{-0.445}$ & $\phn$0.9 & $\phn$0.7  \\
989$^{*}$ & 3.1$\pm$0.3 & 0.2$\pm$0.1 &  10.7$^{+0.2}_{-0.3}$ & \phn9.1$\pm$0.3 & 0.979$^{+0.014}_{-0.049}$ & $\phn$1.6 & $\phn$0.1  \\
1078$^{*}$ & 2.9$^{+0.3}_{-0.2}$ & 0.1$\pm$0.1 &  10.8$\pm$0.2 & \phn7.5$^{+0.6}_{-0.4}$ & 1.000$^{+0.000}_{-0.001}$ & $\phn$6.6 & $\phn$0.6  \\
1082 & 2.7$^{+0.0}_{-0.1}$ & 0.4$\pm$0.0 & \phn9.6$^{+0.1}_{-0.2}$ & \phn9.2$\pm$0.0 & 0.707$^{+0.062}_{-0.112}$ & $\phn$6.0 & $\phn$0.0  \\
1172$^{*}$ & 2.0$\pm$0.4 & 1.1$^{+0.3}_{-0.8}$ &  10.4$^{+0.2}_{-0.6}$ &  10.1$^{+0.6}_{-2.3}$ & 0.513$^{+0.485}_{-0.397}$ & $\phn$1.0 & $\phn$0.8  \\
1184 & 3.2$^{+0.2}_{-0.3}$ & 0.2$\pm$0.1 &  10.9$\pm$0.2 & \phn9.0$\pm$0.3 & 0.987$^{+0.009}_{-0.029}$ & $\phn$0.4 & $\phn$0.2  \\
1211$^{*}$ & 2.9$^{+0.4}_{-0.2}$ & 0.4$\pm$0.1 &  11.2$^{+0.3}_{-0.2}$ & \phn9.8$\pm$0.3 & 0.960$^{+0.028}_{-0.081}$ & $\phn$9.5 & $\phn$1.0  \\
1237$^{*}$ & 3.3$^{+0.3}_{-0.4}$ & 0.1$^{+0.2}_{-0.1}$ &  11.0$\pm$0.2 & \phn8.3$^{+0.5}_{-0.4}$ & 0.998$^{+0.001}_{-0.008}$ & $\phn$2.7 & $\phn$0.0  \\
1245 & 2.8$^{+0.3}_{-0.1}$ & 0.3$^{+0.0}_{-0.1}$ & \phn9.8$\pm$0.5 & \phn9.3$^{+0.1}_{-0.2}$ & 0.766$^{+0.182}_{-0.260}$ & $\phn$5.6 & $\phn$1.7  \\
1267 & 3.2$^{+0.2}_{-0.3}$ & 0.4$\pm$0.1 &  11.6$^{+0.2}_{-0.3}$ &  10.6$\pm$0.3 & 0.921$^{+0.054}_{-0.180}$ & $\phn$5.1 & $\phn$0.0  \\
\enddata
\label{tab2}
\tablenotetext{a}{The E(B-V) extinction is applied only to the young component}
\tablenotetext{b}{The model fit mass in the old (SSP) component.}
\tablenotetext{c}{The model fit mass in the young (constant SFR) component.}
\tablenotetext{d}{The fraction by mass of the old component relative
to the total fitted mass.}
\tablenotetext{e}{The best-fit model $\chi_{\nu}^2$ is listed.}
\tablenotetext{f}{The difference between the $\chi^2$ value 
  for the model listed in columns $2$ through $4$ and
  the best $\chi^2$:  \\
  $\Delta \chi^2$ = $\chi_{median}^2 -$
  $\chi_{best}^2$ (also see Section $3.1$).}
\tablecomments{Columns $2$ through $4$ list the median values and
  the $16\%$ and $82\%$ intervals for the fitted parameters.  
  Objects with unblended IRAC photometry are marked with
  an asterisk symbol.  Objects 30, 88, 1237 and 1267 are detected in x-rays.}
\end{deluxetable*}

\section{SED fitting: mass estimates from a simple photometric
  redshift code}

\subsection{Fitting method}

To study the properties of our selected objects, we fit
stellar-population synthesis models to the photometry and derive
redshifts and stellar masses.  We model the galaxies as the sum of two
stellar components, young and old, and consider the effects of
reddening.  Because the star-formation history is constrained
generally poorly by SED fitting, and the stellar population age and
dust extinction have strong degeneracies (see, e.g., Papovich et al.\
2001), we consider only non-negative linear superpositions of these
two models.  The combinations of these two models encompass the range
of plausible star-formation histories, and in particular, as we
discuss below, they allow us to derive a stringent upper limit on the
amount of stellar mass in old stellar populations.

We therefore have four model parameters: the redshift, dust
extinction, mass in the young stellar component, and the mass in the
older stellar component.  We use the \citet{bc03} population synthesis
code to generate the two stellar population components for our SED
fitting routine: a $100$ Myr constant star formation (composite
stellar population: CSP) model and a maximally old passively evolving
(simple stellar population: SSP) model formed at z = 7, both generated
at solar metallicity and with a Salpeter initial mass function (IMF).
Our maximally old model age is a function of redshift: the age at a
given redshift is equal to the age of the universe at that redshift
minus 0.75~Gyr, the age of the universe at z = 7.  We attenuate the
CSP model using the \citet{dc00} dust law; the old model is not
attenuated.  At each $z$ and $E(B-V)$, we compute model photometry by
convolving the \citet{bc03} spectra with the $8$ filter response
curves \citep[e.g.,][]{cp01}, taking in to account the HI opacity of
the intergalactic medium (IGM) along the line of sight \citep{pm95},
according the models described in \citet{fan99}.

We fit the object fluxes to this four-parameter model assuming that
the errors in the fluxes are symmetric and Gaussian.  In addition to
the photometric uncertainties on the magnitude of each bandpass, we
add a systematic uncertainty proportional to the flux density of
$\sigma_\mathrm{sys} / f_\nu = 0.04$, 0.08, and 0.15 for the ACS,
NICMOS, and IRAC photometry, respectively.  These errors represent
systematic uncertainties in our matched-band photometry, as well as
the fact that our model does not sample the full range of possible
parameters infinitesimally.  These systematic errors are added in
quadrature to either the fitted errors, for the ACS photometry, or the
catalog errors for the NICMOS and IRAC photometry.  We use the
resulting $\chi^2$ statistic to produce the likelihood $\calL \propto
\exp(-\chi^2/2)$ and we take the likelihood to be the probability
distribution for the 4 parameters given the observed fluxes.

\begin{deluxetable*}{lccccccc}
\tabletypesize{\scriptsize}
\tablecaption{Best-fit parameters for objects with IRAC photometry}
\tablewidth{0pt}
\tablehead{
\colhead{NIC ID} 
& \colhead{z$_{phot}$} 
& \colhead{E(B-V)$^{a}$} 
& \colhead{log M$_{o}$/M$_\sun$$^{b}$} 
& \colhead{log M$_{y}$/M$_\sun$$^{c}$} 
& \colhead{f$_{old}$$^{d}$} 
& \colhead{$\chi_{\nu,best}^{2}$$^{e}$}
& \colhead{$\Delta \chi^{2}$$^{f}$}
}
\startdata
30$^{**}$ & 1.8$^{+0.2}_{-0.1}$ & 0.3$^{+0.1}_{-0.0}$ &  11.5$\pm$0.1 & \phn9.7$^{+0.5}_{-0.1}$ & 0.984$^{+0.002}_{-0.026}$ & $\phn$9.2 & $\phn$0.0  \\
88$^{**}$ & 2.2$^{+0.2}_{-0.1}$ & 0.7$^{+0.0}_{-0.1}$ &  10.8$\pm$0.2 &  10.3$^{+0.1}_{-0.2}$ & 0.718$^{+0.158}_{-0.071}$ & $\phn$0.7 & $\phn$0.0  \\
111 & 1.5$^{+0.1}_{-0.0}$ & 0.4$\pm$0.0 & \phn9.6$\pm$0.1 & \phn8.5$^{+0.1}_{-0.0}$ & 0.914$^{+0.010}_{-0.015}$ & 11.3 & $\phn$0.0  \\
120$^{**}$ & 3.4$^{+0.1}_{-0.2}$ & 0.2$\pm$0.1 &  11.1$\pm$0.1 & \phn8.9$\pm$0.4 & 0.994$^{+0.004}_{-0.012}$ & $\phn$2.0 & $\phn$0.2  \\
219$^{\bigstar}$ & 2.8$^{+0.2}_{-0.0}$ & 0.3$^{+0.0}_{-0.1}$ &  11.1$\pm$0.1 & \phn9.3$^{+0.0}_{-0.3}$ & 0.982$^{+0.011}_{-0.002}$ & $\phn$2.2 & $\phn$0.0  \\
223$^{\bigstar}$ & 3.1$\pm$0.1 & 0.2$^{+0.1}_{-0.0}$ &  11.4$\pm$0.1 & \phn9.1$^{+0.3}_{-0.0}$ & 0.995$^{+0.000}_{-0.007}$ & $\phn$0.7 & $\phn$0.0  \\
269$^{\bigstar}$ & 2.7$^{+0.1}_{-0.0}$ & 0.2$\pm$0.0 &  10.4$\pm$0.1 & \phn9.2$\pm$0.0 & 0.944$^{+0.007}_{-0.008}$ & $\phn$1.9 & $\phn$0.0  \\
319$^{**}$ & 1.2$^{+0.0}_{-0.1}$ & 0.9$^{+0.1}_{-0.0}$ &  10.0$^{+0.2}_{-0.4}$ &  10.3$\pm$0.1 & 0.344$^{+0.129}_{-0.156}$ & $\phn$1.3 & $\phn$0.0  \\
625$^{\bigstar}$ & 2.7$\pm$0.1 & 0.4$^{+0.0}_{-0.1}$ &  10.6$\pm$0.1 & \phn9.4$^{+0.1}_{-0.3}$ & 0.937$^{+0.039}_{-0.015}$ & $\phn$3.3 & $\phn$0.0  \\
706 & 1.3$^{+0.2}_{-0.0}$ & 0.6$\pm$0.0 &  10.3$^{+0.2}_{-0.3}$ & \phn9.9$\pm$0.1 & 0.703$^{+0.095}_{-0.137}$ & $\phn$4.2 & $\phn$0.0  \\
921 & 1.9$\pm$0.1 & 0.5$\pm$0.1 &  10.9$\pm$0.1 & \phn9.5$\pm$0.3 & 0.966$^{+0.019}_{-0.045}$ & $\phn$1.6 & $\phn$0.0  \\
989$^{\bigstar}$ & 2.9$^{+0.1}_{-0.2}$ & 0.3$^{+0.0}_{-0.1}$ &  10.5$^{+0.2}_{-0.1}$ & \phn9.3$^{+0.1}_{-0.3}$ & 0.935$^{+0.042}_{-0.019}$ & $\phn$1.3 & $\phn$1.5  \\
1078 & 2.9$^{+0.7}_{-0.3}$ & 1.0$^{+0.0}_{-1.0}$ &  10.6$^{+0.6}_{-0.3}$ &  10.7$^{+0.1}_{-3.3}$ & 0.386$^{+0.614}_{-0.123}$ & $\phn$5.8 & $\phn$2.4  \\
1172 & 2.1$^{+0.2}_{-0.3}$ & 1.1$^{+0.3}_{-0.1}$ &  10.4$^{+0.1}_{-0.2}$ &  10.1$^{+0.1}_{-0.2}$ & 0.661$^{+0.078}_{-0.072}$ & $\phn$1.0 & $\phn$0.3  \\
1211$^{\bigstar}$ & 2.7$\pm$0.1 & 0.4$^{+0.1}_{-0.0}$ &  11.0$\pm$0.1 & \phn9.9$\pm$0.2 & 0.949$^{+0.013}_{-0.073}$ & $\phn$5.0 & $\phn$0.5  \\
1237$^{**}$ & 3.6$^{+0.2}_{-0.0}$ & 0.0$^{+0.1}_{-0.0}$ &  11.2$^{+0.1}_{-0.0}$ & \phn8.0$^{+0.4}_{-0.1}$ & 0.999$^{+0.000}_{-0.001}$ & $\phn$1.9 & $\phn$0.0  \\
\enddata
\label{tab3}
\tablenotetext{a}{The E(B-V) extinction is applied only to the young component}
\tablenotetext{b}{The model fit mass in the old (SSP) component.}
\tablenotetext{c}{The model fit mass in the young (constant SFR) component.}
\tablenotetext{d}{The fraction by mass of the old component relative
to the total fitted mass.}
\tablenotetext{e}{The best-fit model $\chi_{\nu}^2$ is listed.}
\tablenotetext{f}{The difference between the $\chi^2$ value 
  for the model listed in columns $2$ through $4$ and
  the best $\chi^2$:  \\
  $\Delta \chi^2$ = $\chi_{median}^2 -$
  $\chi_{best}^2$ (also see Section $3.1$).}
\tablecomments{Columns $2$ through $4$ list the median values and
  the $16\%$ and $82\%$ intervals for the fitted parameters.    
  The six galaxies with broad-band SEDs consistent with distant
  evolved stellar populations are marked with a star symbol.  
  Objects detected at $24\micron$ are indicated with
  a double-asterisk symbol.  Objects 30, 88, and 1237 are detected in x-rays.}
\end{deluxetable*}

We derive mean values and variances for the 4 parameters (and
functions thereof) by integrating over the full 4-dimensional
likelihood.  The integrations are done with a hybrid grid plus Monte
Carlo approach.  The $\chi^2$ is quadratic in two parameters (the old
and young masses, $M_o$ and $M_y$) and more complicated in the other
two (redshift and reddening).  We construct a grid in redshift and
reddening over the range $0<z<5$ and $0<E(B-V)<1.5$.  For each point,
we use linear methods to find the best fit and Gaussian covariance
matrix for the masses.  We generate Gaussian-distributed points given
this covariance region and weight points by $\exp(-\chi^2_{\rm
best}/2)$ for the best-fit $\chi^2$, dropping any points that have
negative masses.  The concatenation of all of the points is a Monte
Carlo of the full 4-dimensional likelihood, and any function can be
computed by a simple weighted sum of the function evaluated at all of
the points.  Similarly, we can construct quantiles by sorting the
points according to a given function evaluation and cumulating the
weights.  By using a grid in the non-quadratic directions, we ensure
that multiple minima are found and correctly weighted.
  
The derived physical parameters are presented in table~\ref{tab2} and
table~\ref{tab3}.  In these tables we define $f_\mathrm{old}$ to be
the ratio of the mass in the passively evolving component, $M_\mathrm
o$ to the total mass, $M_\mathrm o + M_\mathrm y$, where $M_\mathrm y$
is the mass in the star-forming component,
\begin{equation} 
f_\mathrm{old} = M_\mathrm o / (M_\mathrm o + M_\mathrm y).
\end{equation}
We characterize the probability distribution function for the various
parameters by the median and the $16\%$ and $84\%$ intervals
(equivalent to the $1\sigma$ intervals for a Gaussian distribution).
However, in cases with multiple minima these simple quantiles will not
fully represent the distribution.

\begin{figure*}[t]
  \begin{center}
    \scalebox{0.43}{{\includegraphics{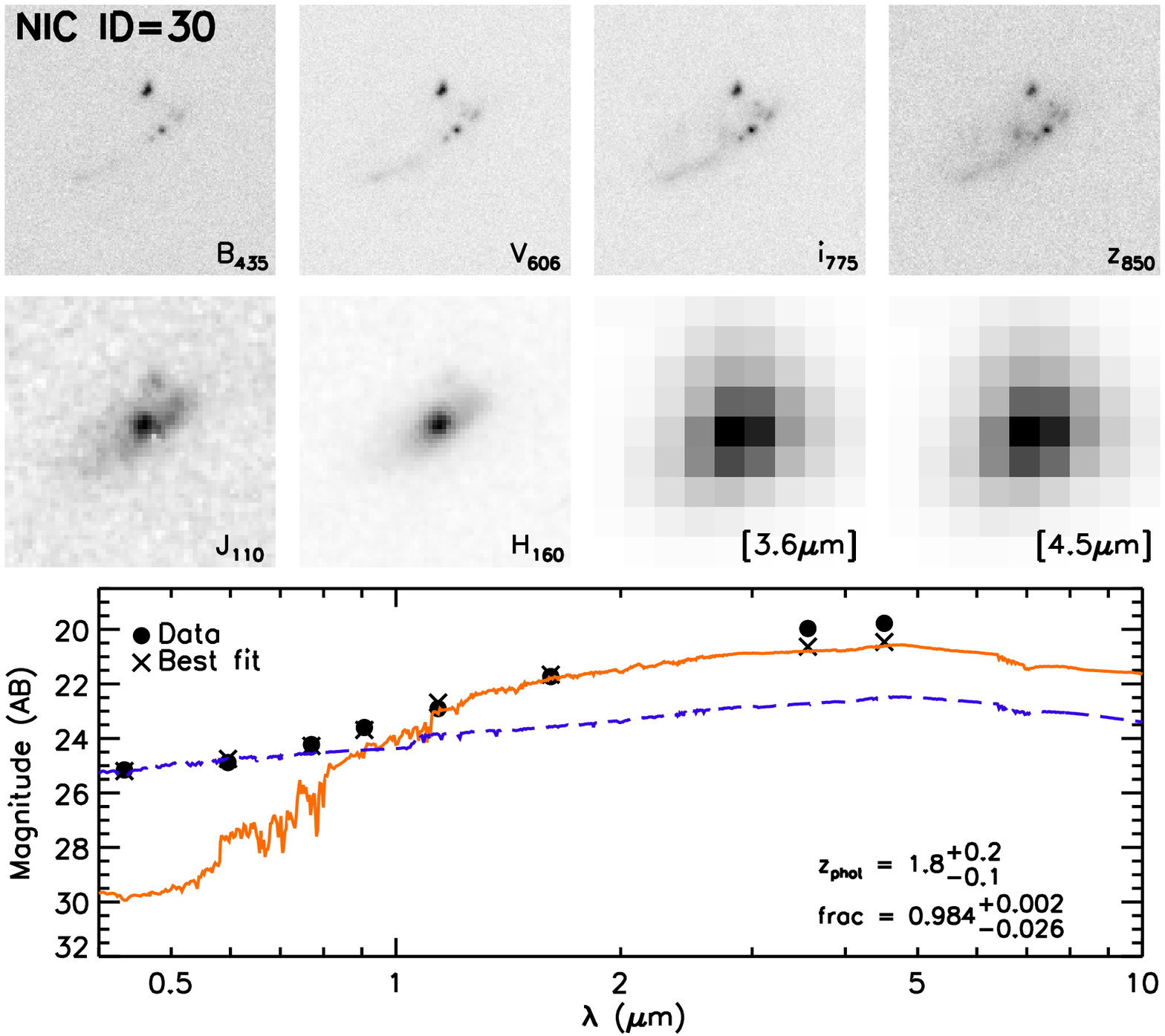}}
      {\includegraphics{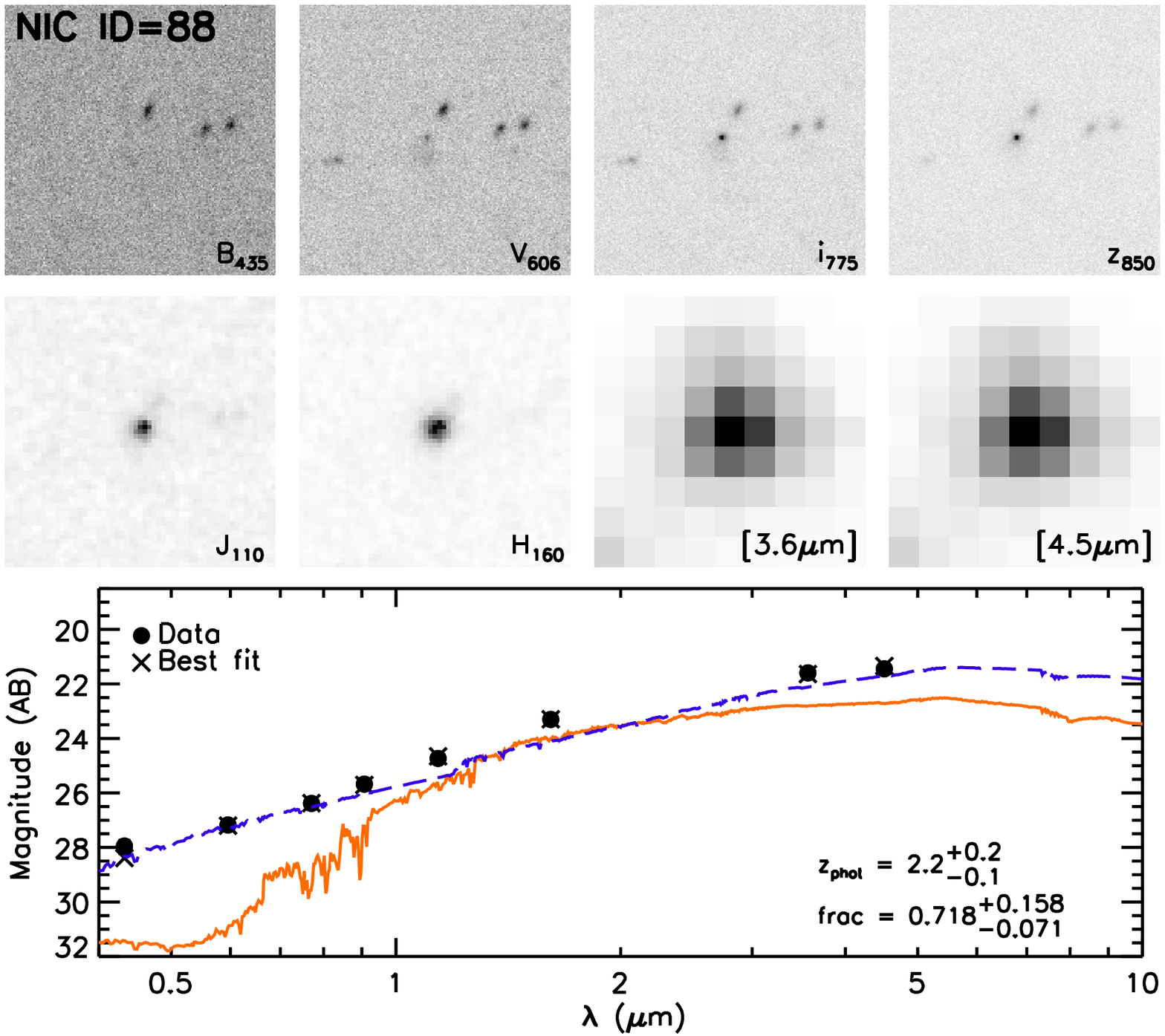}}}
    \scalebox{0.43}{{\includegraphics{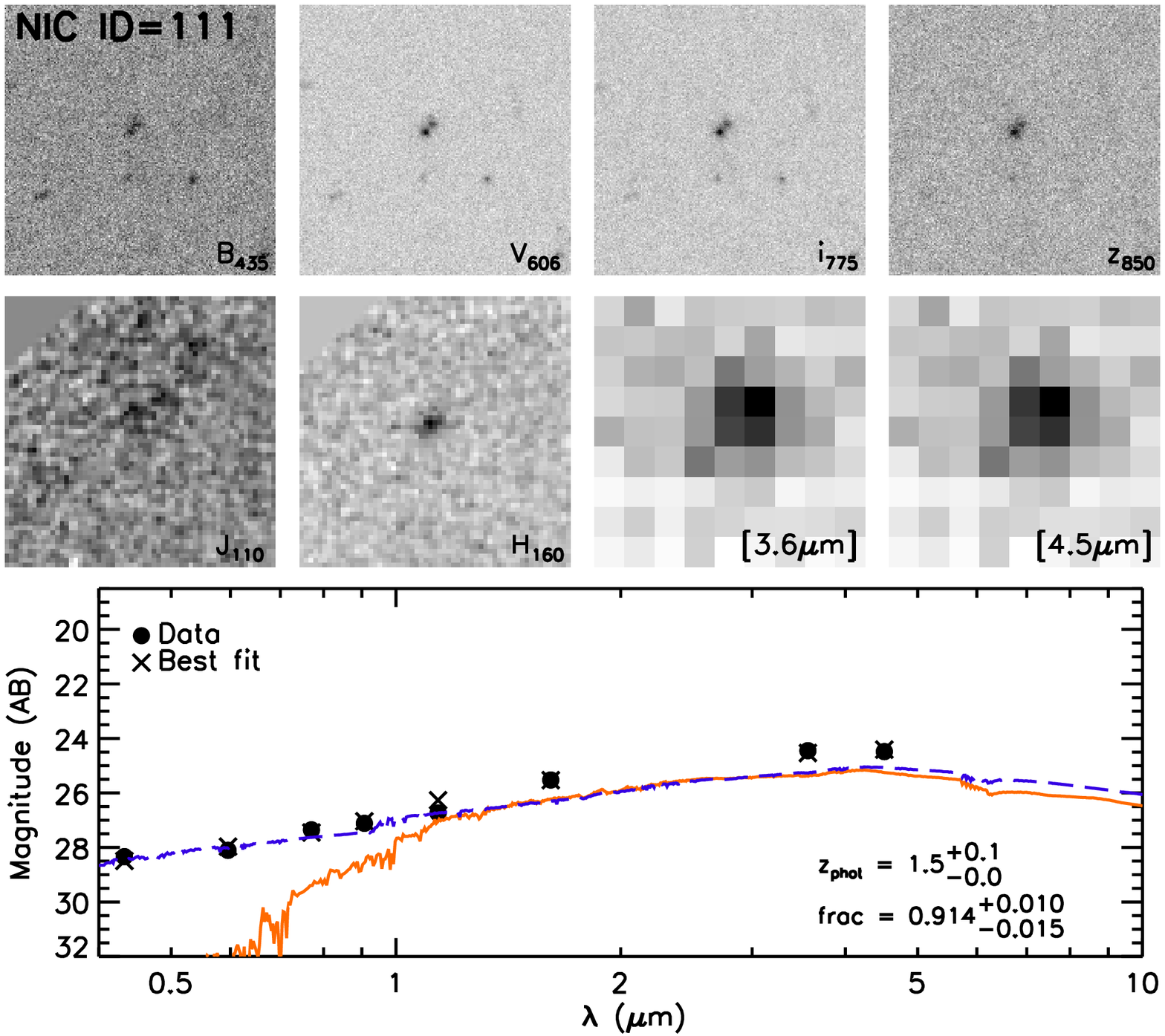}}
      {\includegraphics{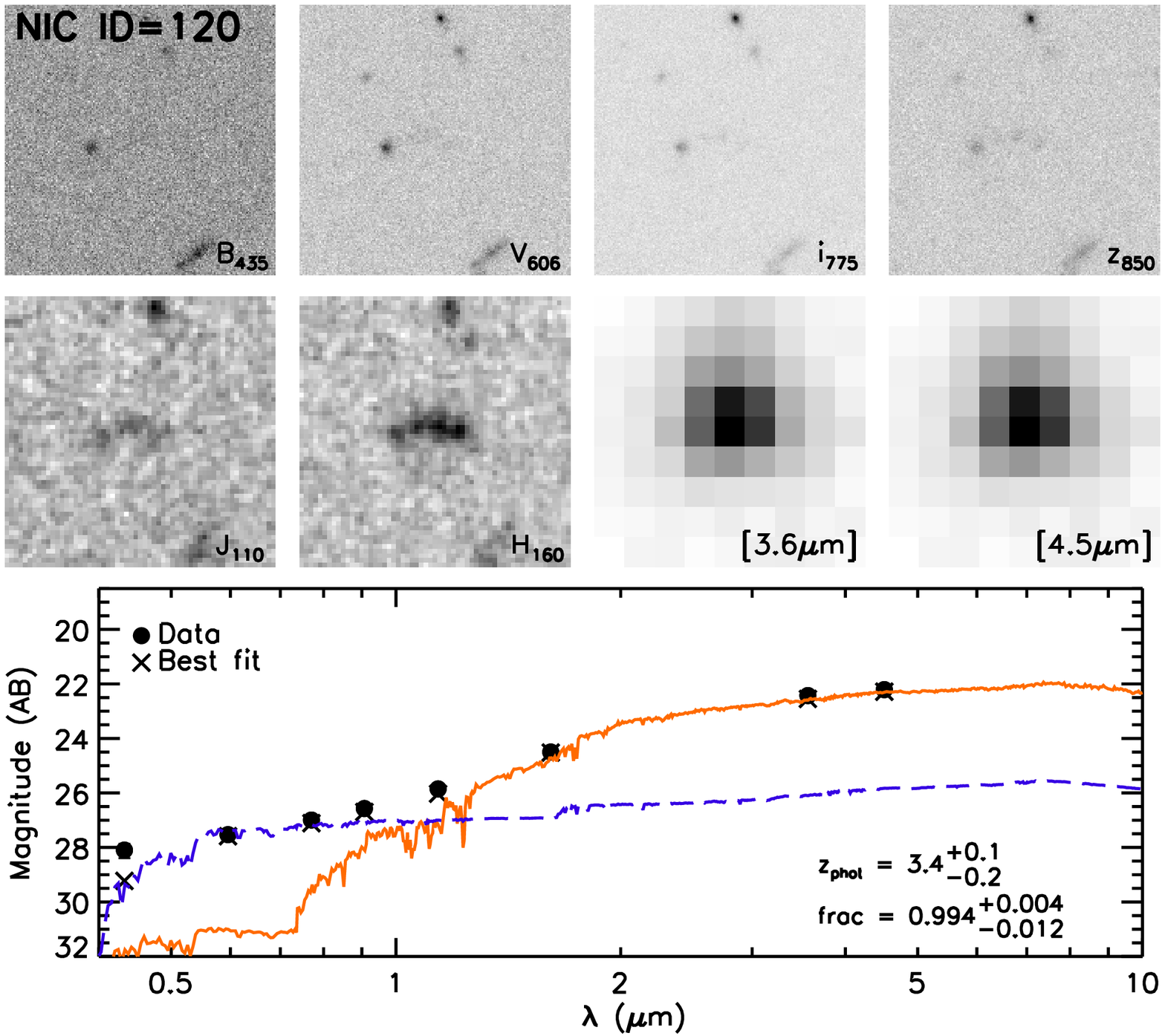}}}
    \caption{ Gallery of $J_{110} - H_{160} > 1.0$ and $H_{160} <
    26.0$ selected objects; we show imaging, broad-band SED's and the
    best fit model for each object.  The
    $B_{435}V_{606}i_{775}z_{850}J_{110}H_{160}[3.6\mu m][4.6\mu m]$
    postage stamps are $5\farcs4$ on a side and oriented such that
    North is up and East is to the left.  The
    $B_{435}V_{606}i_{775}z_{850}J_{110}H_{160}$ stamps are plotted at
    constant $f_{\nu}$ per arcsec and are displayed on a log scale.
    The IRAC imaging is also displayed at constant $f_{\nu}$ per
    arcsec on a log scale but on a different relative scale than the
    $B_{435}V_{606}i_{775}z_{850}J_{110}H_{160}$ imaging.  The
    broad-band SED's are shown in the lower panels with solid circles,
    $1\sigma$ error bars are indicated.  The crosses indicate the best
    fit model.  The dashed line indicates the fraction, by mass, of
    light that is attributed by the best fit to the young model;
    similarly, the solid line indicates the fraction attributed to the
    old model.\label{fig2}}
  \end{center}
\end{figure*}

\subsection{Fitting results}
  
We apply this SED fitting method to derive physical parameters for our
objects.  We present the results of the ACS and NICMOS photometry
fitting analysis in detail in Table~\ref{tab2} and summarize them here
as follows.

The fits to 11 objects in our sample favor models with relatively
lower redshifts ($z_\mathrm{phot} < 2.5)$ and lower stellar mass
fractions in old stellar populations ($f_\mathrm{old} < 0.9$): NIC IDs
$86$, $88$, $99$, $215$, $234$, $319$, $535$, $683$, $902$, $921$, and
$1172$.  We note that on average this set of galaxies has more dust,
with a mean fitted $E(B - V) \sim 0.8$, than the entire sample, which
has a mean fitted $E(B - V) \sim 0.4$.  The fits to 7 objects favor
models with lower redshifts and higher mass fractions in older stars
($z_\mathrm{phot} < 2.5$ and $f_\mathrm{old} > 0.9$): $30$, $111$,
$120$, $245$, $281$, $706$, and $935$.  For objects 1082 and 1245, the
models favor a solution with high redshift ($z_\mathrm{phot} > 2.5$)
and a relatively low fraction of mass in old stars ($f_\mathrm{old} <
0.9$).  For 12 objects the models favor solutions with high redshift
($z_\mathrm{phot} > 2.5$) and a relatively high fraction of stars in
the old stellar population ($f_\mathrm{old} > 0.9$): 66, 219, 223,
269, 478, 625, 989, 1078, 1184, 1211, 1237 and 1267.

We find that when we include the IRAC [3.6\micron] and [4.5\micron]
data in the model fitting, 12 objects (of the 16 with unblended IRAC
photometry) have consistent results compared to the model fits without
the IRAC data.  Of these 12 objects, $6$ objects are fitted with
high-redshift model galaxies with only small amounts of rest-frame UV
light associated with a young stellar population or on-going star
formation (for this set $z_\mathrm{phot} \geq 2.7$ and $f_\mathrm{old}
> 0.93$).  The fitted model parameters for these $16$ objects are
listed in Table~\ref{tab3}.  Here we discuss the particularities of
these objects and their corresponding fits.  We will not individually
discuss the objects without IRAC photometry.  However, we note that
object $1267$, is detected in the X-ray data, with a $0.5 - 2.0$ keV
flux of $f_x = 7.5 \times 10^{-16}$ ergs cm$^{-2}$ s$^{-1}$.

\begin{figure*}[t]
  \begin{center}
    \scalebox{0.43}{{\includegraphics{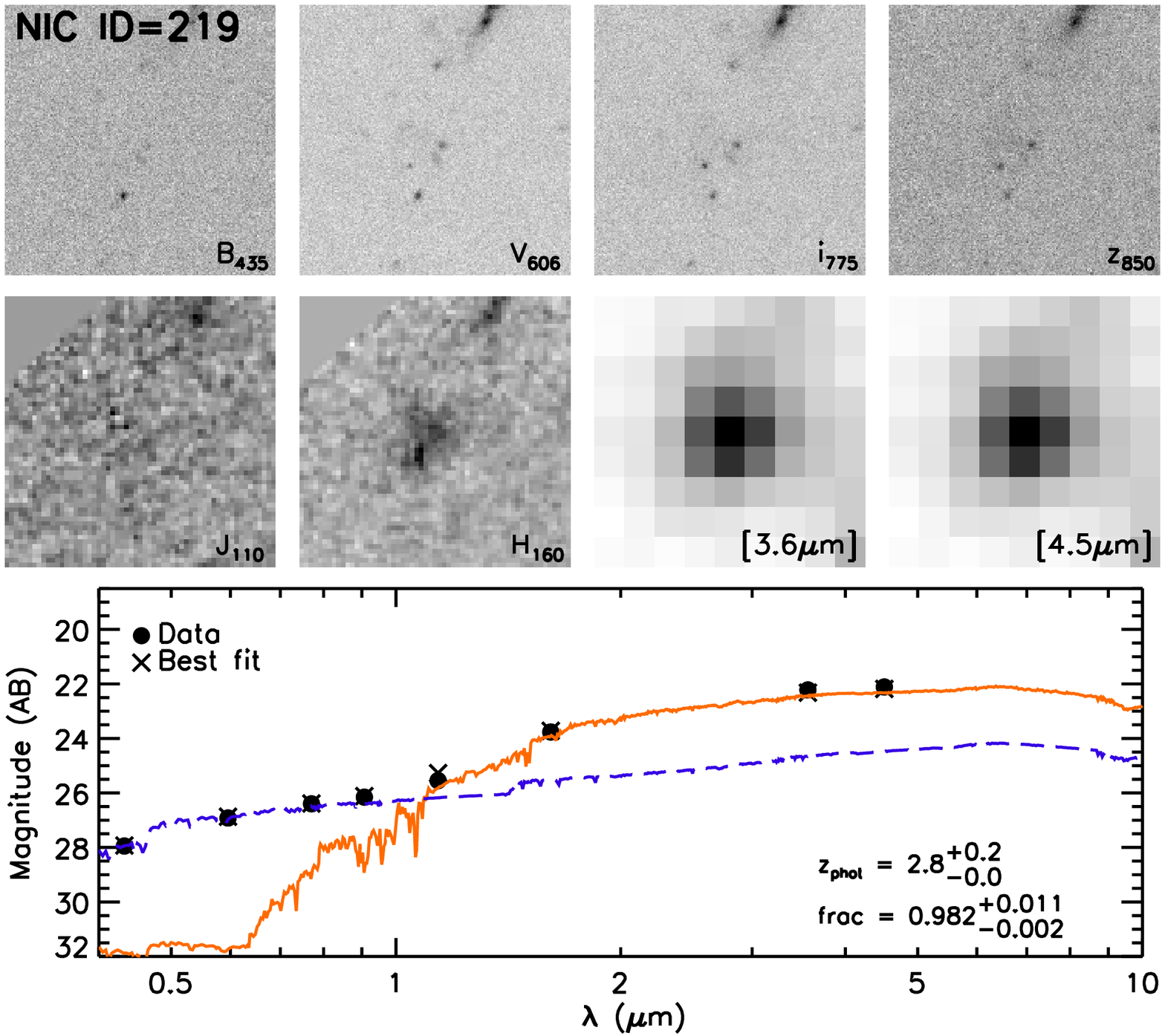}}
      {\includegraphics{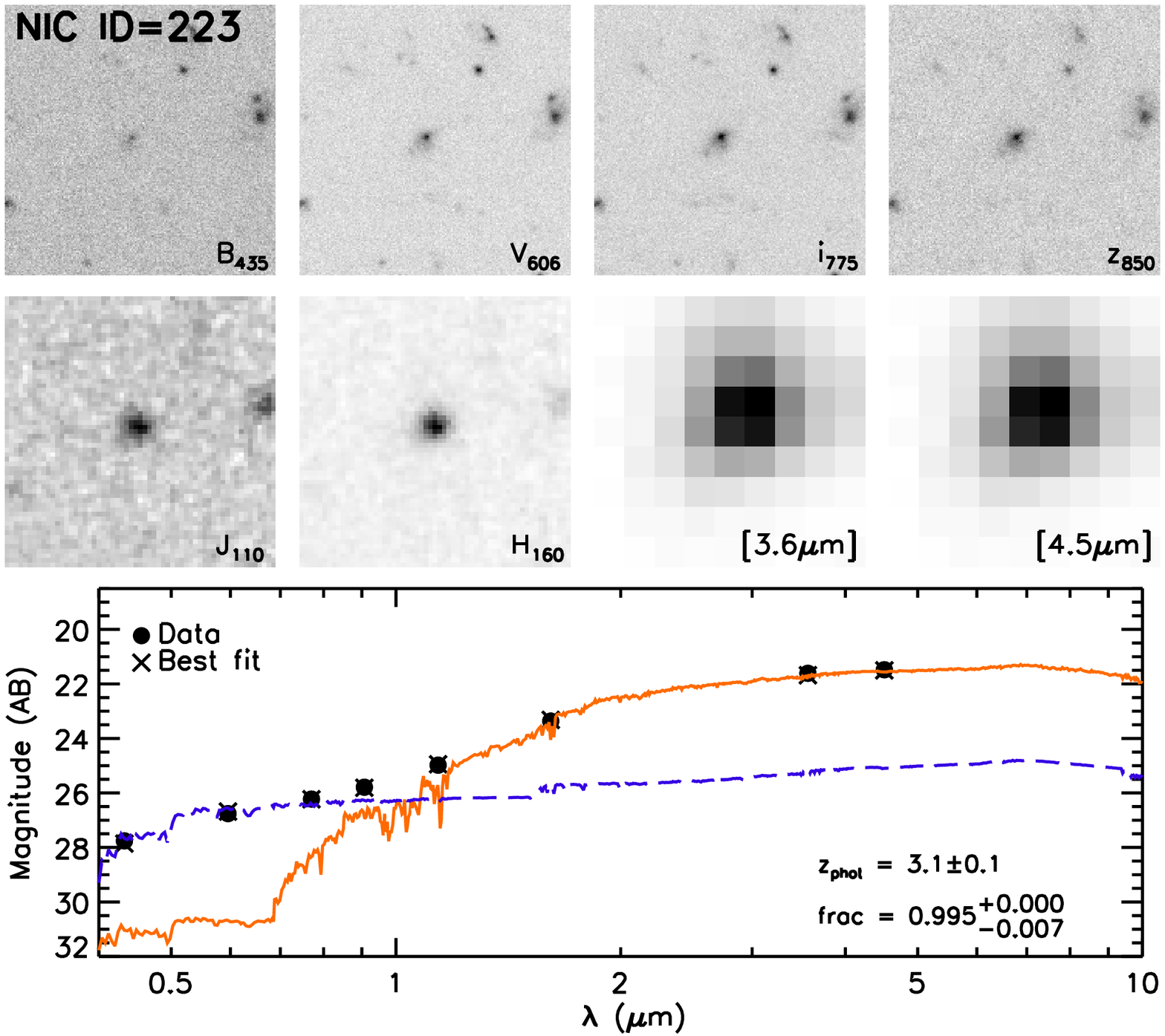}}}
    \scalebox{0.43}{{\includegraphics{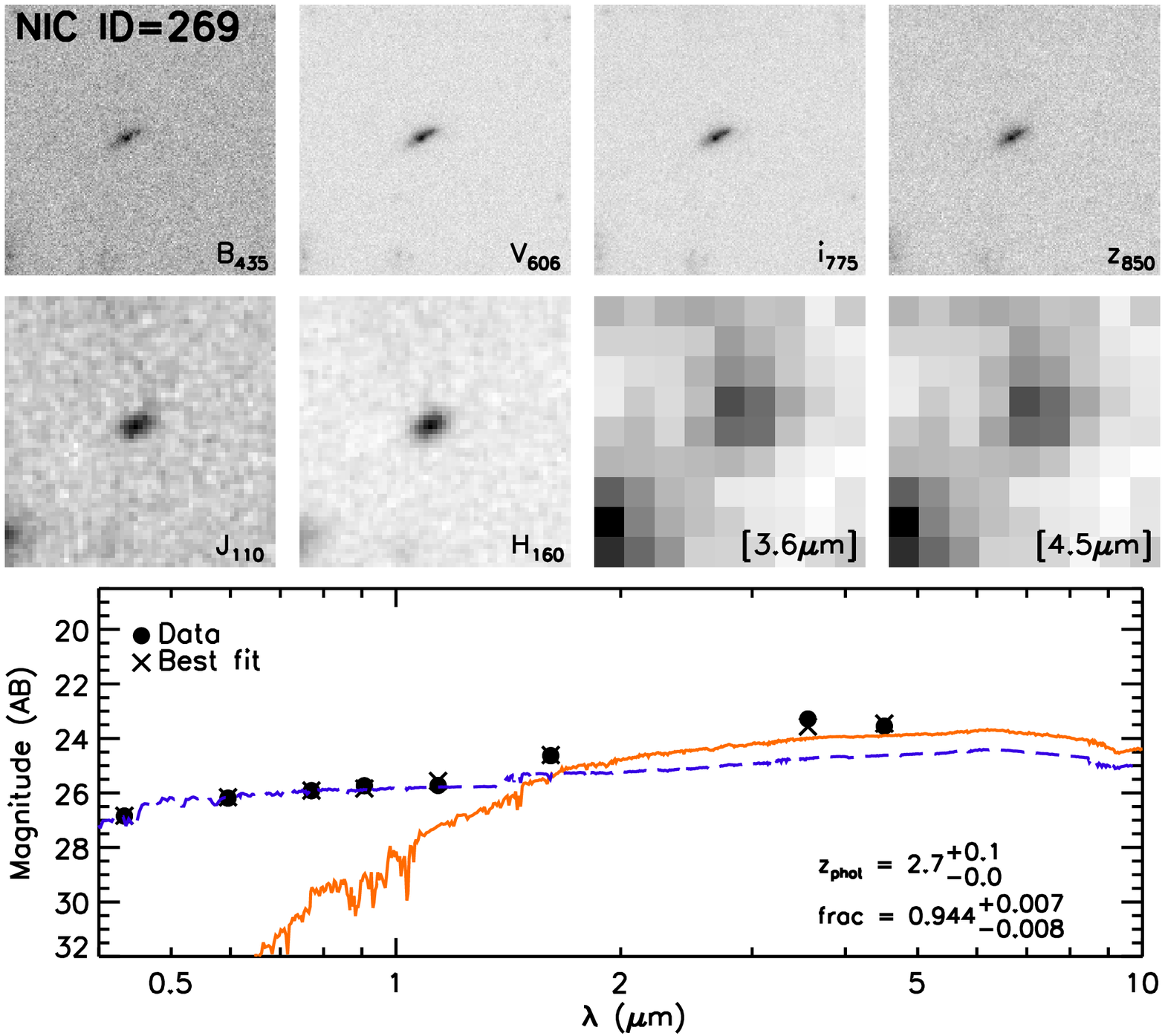}}
      {\includegraphics{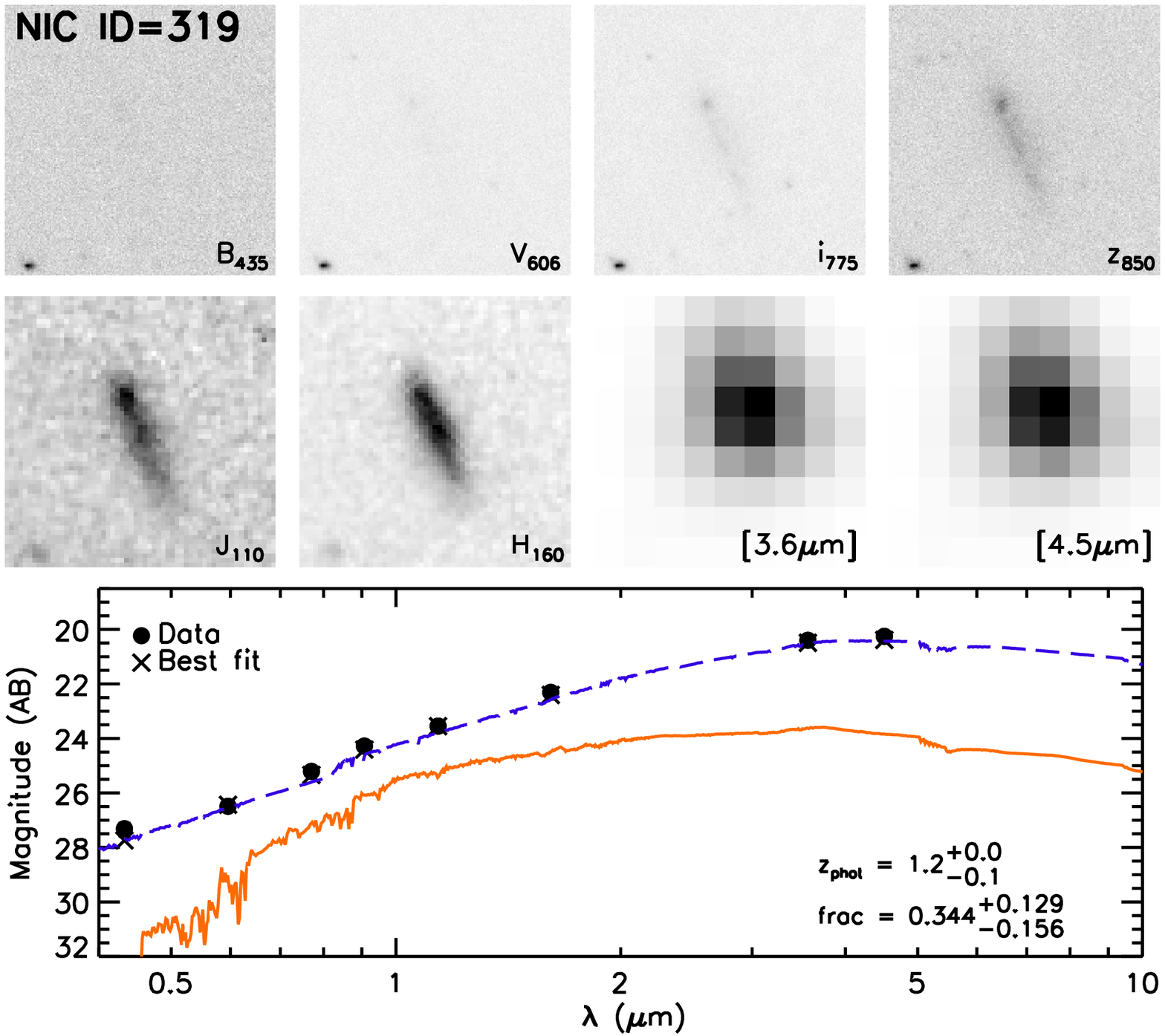}}}
    \caption{Same as figure~\ref{fig2}.\label{fig3}}
  \end{center}
\end{figure*}

{\bf \noindent NIC ID $= 30$:} The difference in the fitted parameters
when we include the IRAC data is significant: $\Delta z_{phot} \sim
0.4$.  As expected, the masses change as well.  However, both values
for $f_\mathrm{old}$ remain high.  The best-fit $\chi^2_\nu$ is very
large for the IRAC fit ($\chi^2_\nu = 9.2$).  This is due to the large
amount of IRAC flux that our models inadequately characterize (see
figure~\ref{fig2}); this excess is also driving the selection of a
higher redshift and mass.  \citet{toft05} derive a photometric
redshift of $z = 2.2$, roughly consistent with our IRAC fit redshift
$z = 1.8^{+0.2}_{-0.1}$.  However, this object is detected with MIPS
at $24\micron$ with $f_\nu(24\micron) = 190$~\ujy, and it has an
X--ray data detection of $f_X(0.5-2\mathrm{keV}) = 8.4 \times
10^{-16}$ ergs cm$^{-2}$ s$^{-1}$.  Based on this data it is likely
that this object likely has an AGN; therefore we consider the lower
redshift fit more plausible.  Furthermore, it is possible that the AGN
contributes to the broadband photometry, and our model fits may not be
valid for this object.

{\bf \noindent NIC ID $= 88$:} The two fits, with and without the IRAC
data, are consistent, with $z_\mathrm{phot} \sim 2.2$ and
$f_\mathrm{old} \sim 0.8$.  It is detected at $24\micron$ with
$f_\nu(24\micron) = 70$~\ujy, although inspection of the $24\micron$
image shows that the object is blended (with a PSF FWHM $\simeq
6$\arcsec).  This object also has a $0.5 - 2.0$ keV x-ray flux $f_X =
8.8 \times 10^{-16}$ ergs cm$^{-2}$ s$^{-1}$ \citep{alex03}.  In
conjunction these data indicate that this object is likely host to an
AGN, a conclusion that is supported by the fact that this object has
red IRAC $[5.8\micron] - [8.0\micron]$ colors and that it has a
compact morphology in the HST images (see figure~\ref{fig2}).  As with
NIC ID 30, the presence of an AGN may imply that the models are
inadequate.

{\bf \noindent NIC ID $= 111$:} The best-fit model, derived including
the IRAC data, is lower redshift older stellar population:
$z_\mathrm{phot} = 1.5$ and $f_\mathrm{old} = 0.914$.  The model
derived from the ACS and NICMOS data alone have a similar
$f_\mathrm{old}$, but the redshift is much lower, $z = 0.7$.  This
discrepancy is driven by the bright IRAC photometry; our models are
simply not red enough at $z = 0.7$ to reproduce the mid-IR data.  This
indicates that the ACS and NICMOS broad-band photometry alone does not
adequately characterize this object.

{\bf \noindent NIC ID $= 120$:} The fitted photometric redshifts and
masses are consistent with a distant old stellar population dominating
the flux in the rest-frame optical and a younger stellar population
(or on-going star formation) dominating in the rest-frame UV, with
$z_\mathrm{phot} = 3.4$ and $f_\mathrm{old} = 0.994$.  The physical
parameters derived without the IRAC data are {\it not} consistent with
this interpretation, with a much smaller distance ($z = 2.4)$ and a
somewhat lower fraction ($f_\mathrm{old} = 0.948$).  Inspection of the
images reveals that the blue component dominating the rest-frame UV
photometry is unresolved and just off-center (see figure~\ref{fig2}).
Therefore, the possibility that this SED is the result of the
projection of more than one object cannot be ruled out.  Despite the
fact that $[5.8\micron]$ and $[8.0\micron]$ IRAC fluxes are in good
agreement with the distant old population scenario, we do not include
this object in our set of distant and evolved object candidates
because of the discrepancy between the two model fits.  Furthermore,
this object is detected at $24\micron$.

{\bf \noindent NIC ID $= 219$:} Our fitted physical parameters are
consistent with a high redshift galaxy dominated by an old stellar
population ($z_\mathrm{phot} = 2.8$ and $f_\mathrm{old} = 0.98$).  The
fit derived from the dataset including IRAC data is consistent with
this model.  The derived redshift is also consistent with the
photometric redshifts of \citet{chen04}.

\begin{figure*}[t]
  \begin{center}
    \scalebox{0.43}{{\includegraphics{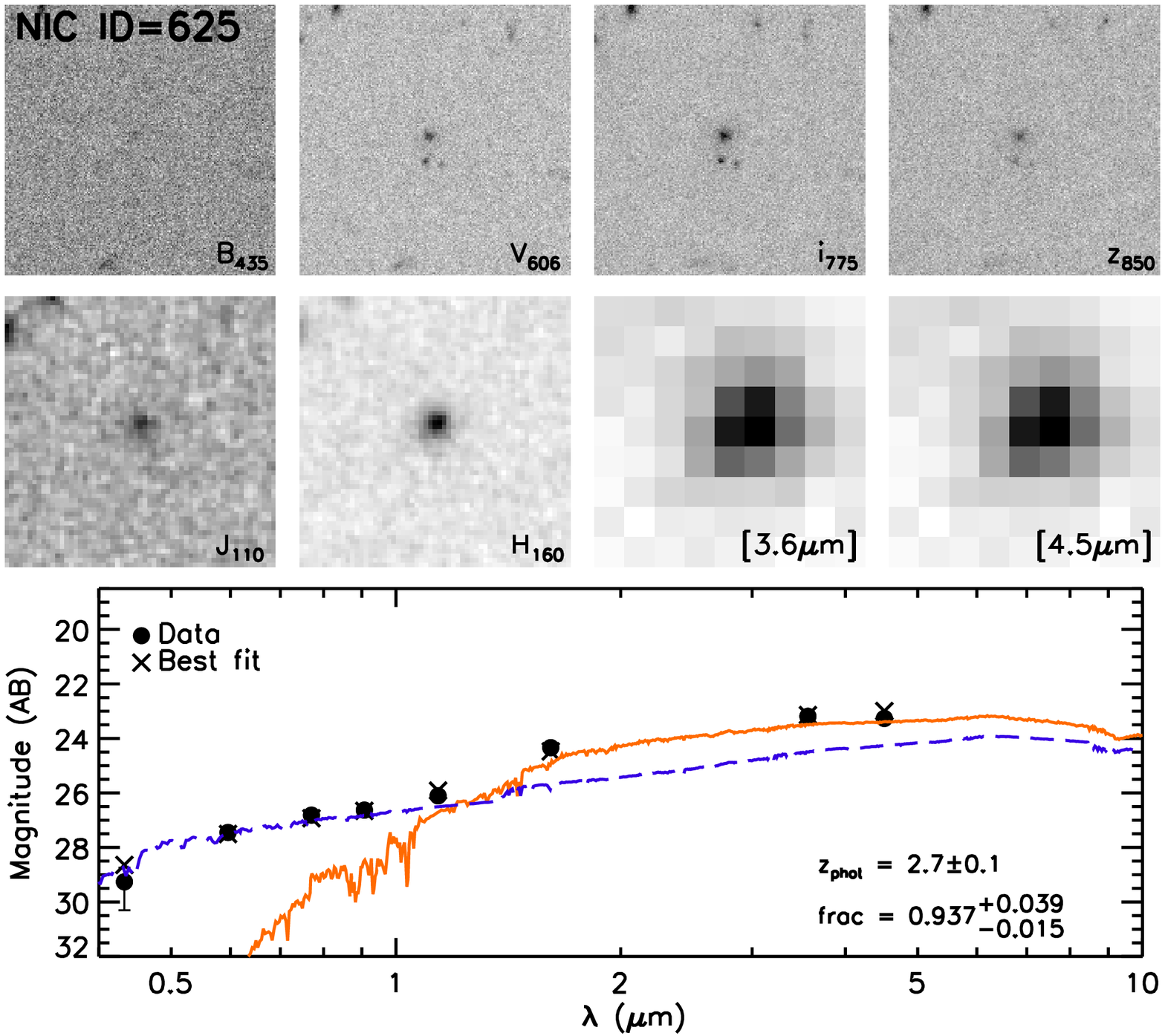}}
      {\includegraphics{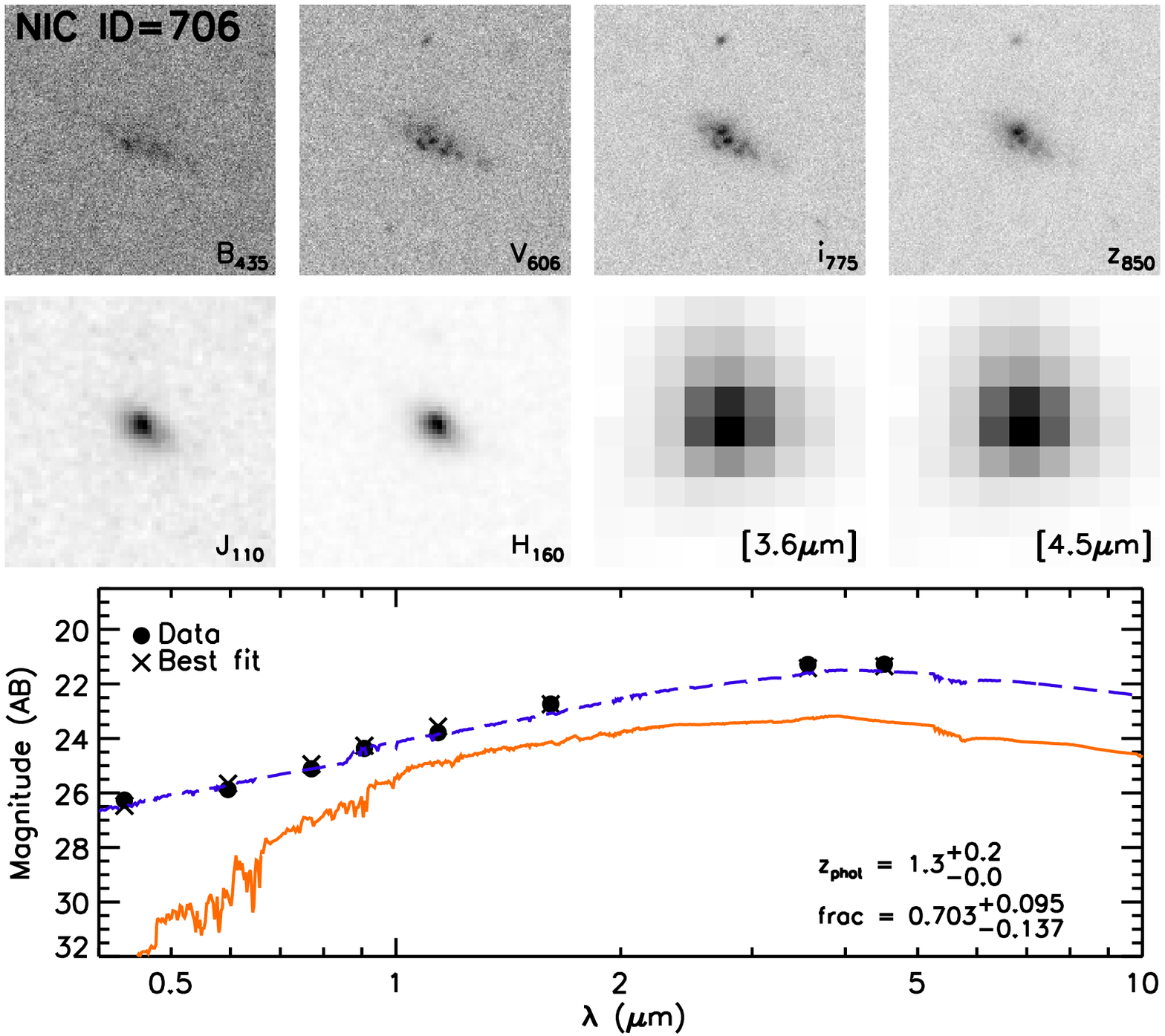}}}
    \scalebox{0.43}{{\includegraphics{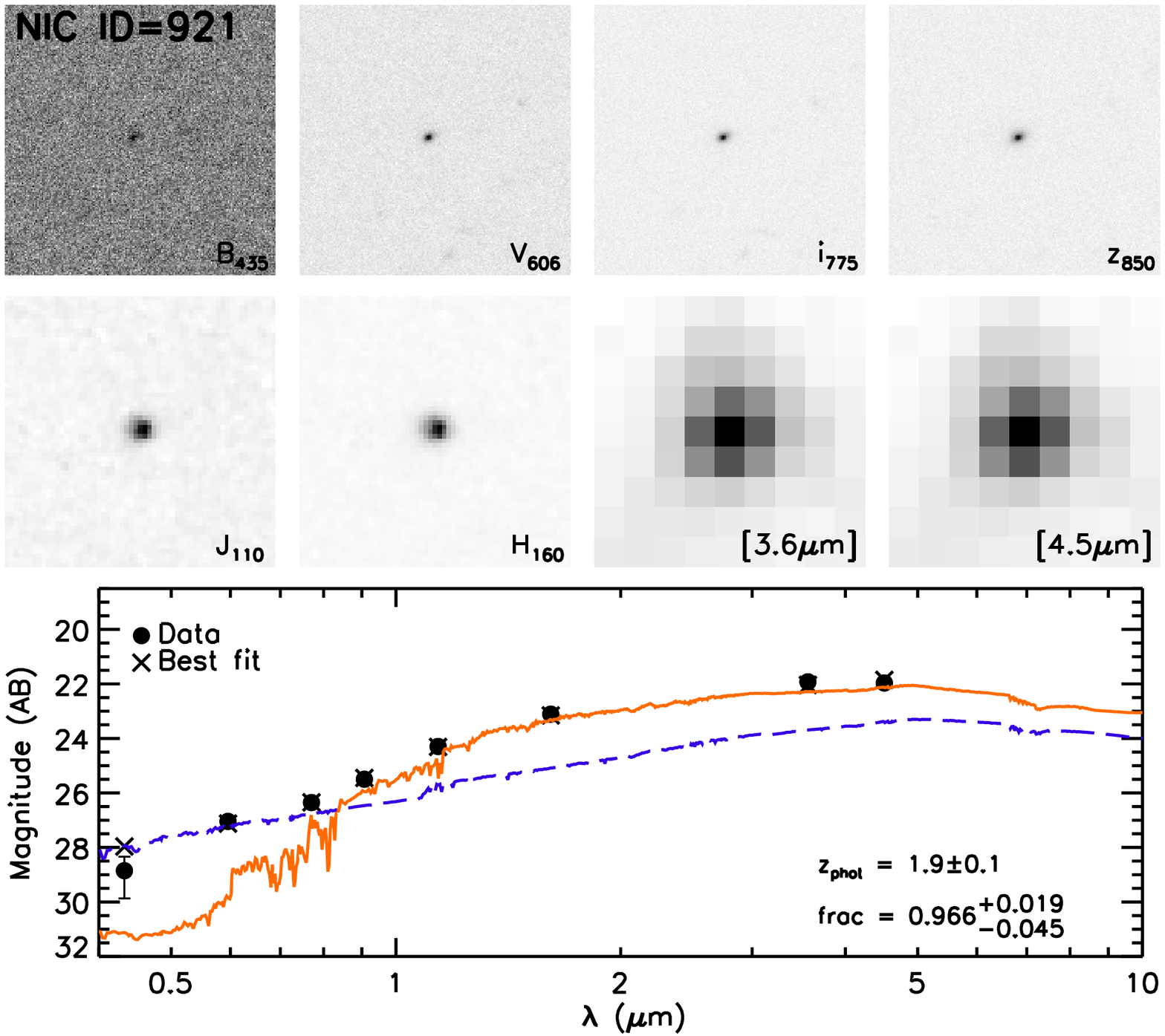}}
      {\includegraphics{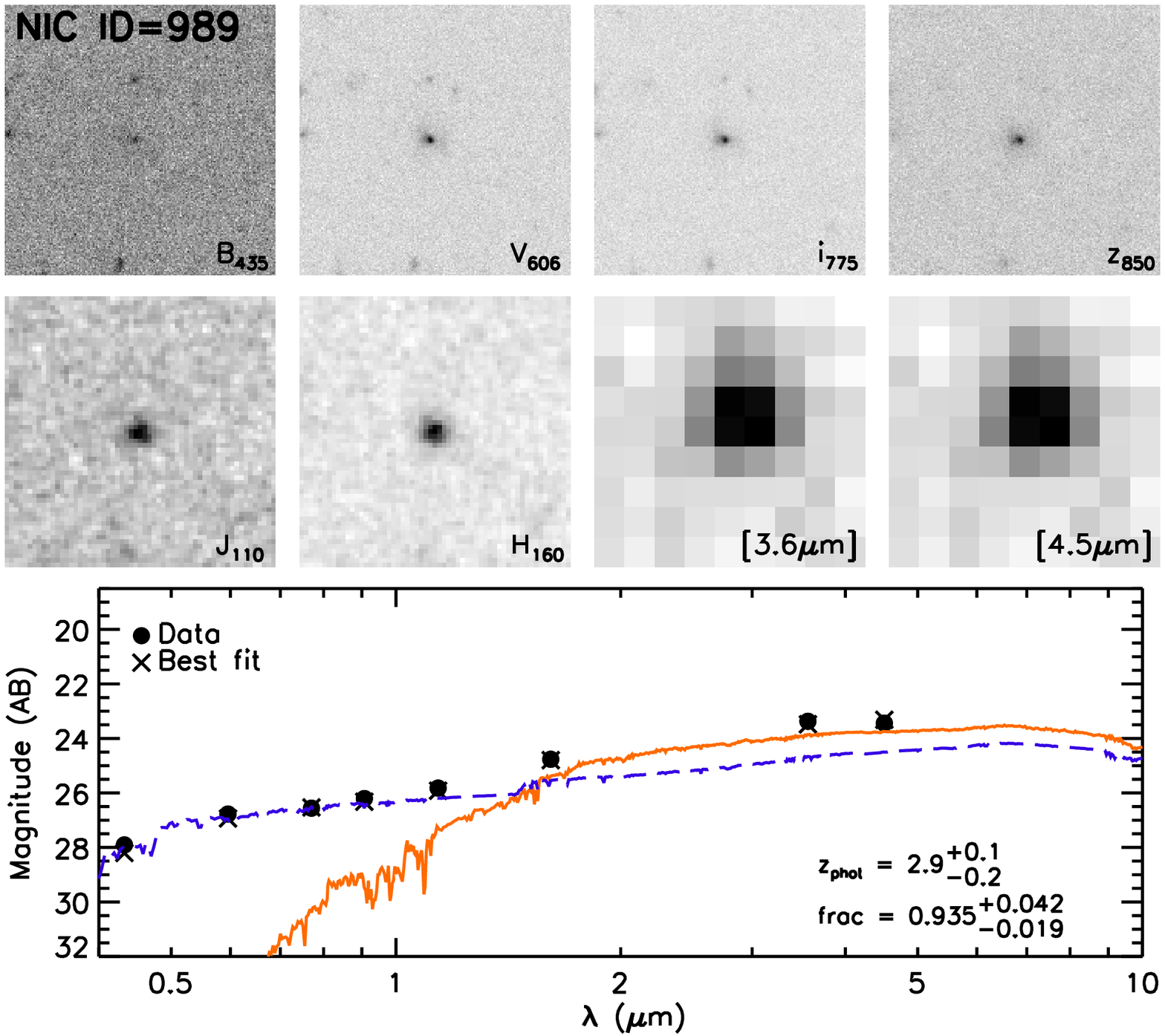}}}
    \caption{Same as figure~\ref{fig2}.\label{fig4}}
  \end{center}
\end{figure*}

{\bf \noindent NIC ID $= 223$:} The fitted model is a distant galaxy
dominated by an old stellar population, with best-fit parameters
$z_\mathrm{phot} = 3.1 \pm 0.1$ and $f_\mathrm{old} = 0.995$,
consistent with the model derived from the ACS and NICMOS data alone.
\citet{chen04} derive a photometric redshift of $z_\mathrm{phot} =
3.11$, in good agreement with our derived redshift. \citet{yan04a}
derive a $z_\mathrm{phot} = 2.9$ and \citet{toft05} derive a
$z_\mathrm{phot} = 3.4$, however, these authors do not quote errors,
making it hard to compare with our model.

{\bf \noindent NIC ID $= 269$:} The fit parameters are consistent with
a distant object dominated by an old stellar population with a small
amount of ongoing star formation, for both fits with and without IRAC
data ($z_\mathrm{phot} = 2.7$ and $f_\mathrm{old} = 0.94$)

{\bf \noindent NIC ID $= 319$:} The most-likely model is a low
redshift dusty starburst: $z_\mathrm{phot} = 1.2$, $E(B-V) = 0.9$ and
$f_\mathrm{old} = 0.34$.  \citet{toft05} derive a $z_\mathrm{phot} =
1.8$ for this object.  This object is detected at $24\micron$ with
$f_\nu(24\micron) = 110$~\uJy, supporting the interpretation that this
object is a lower redshift dust--enshrouded starburst.

{\bf \noindent NIC ID $= 625$:} The most-likely model is consistent
with a distant galaxy dominated by mass by an old stellar population
with a small amount of ongoing star formation ($z_\mathrm{phot} = 2.7
\pm 0.1$ and $f_\mathrm{old} = 0.94$).  The redshift derived without
the IRAC data is higher, although poorly constrained, $z_\mathrm{phot}
= 3.3 \pm 0.4$, and the stellar population remains dominated by an old
population.  \citet{chen04} derive a redshift for this object of
$z_\mathrm{phot} = 3.33$, consistent with our fit excluding the IRAC
data (see table 2).

{\bf \noindent NIC ID $= 706$:} The most-likely model is consistent
with a low-redshift dusty star forming galaxy.  \citet{chen04} reach a
similar conclusion for this object.

\begin{figure*}
  \begin{center}
    \scalebox{0.43}{{\includegraphics{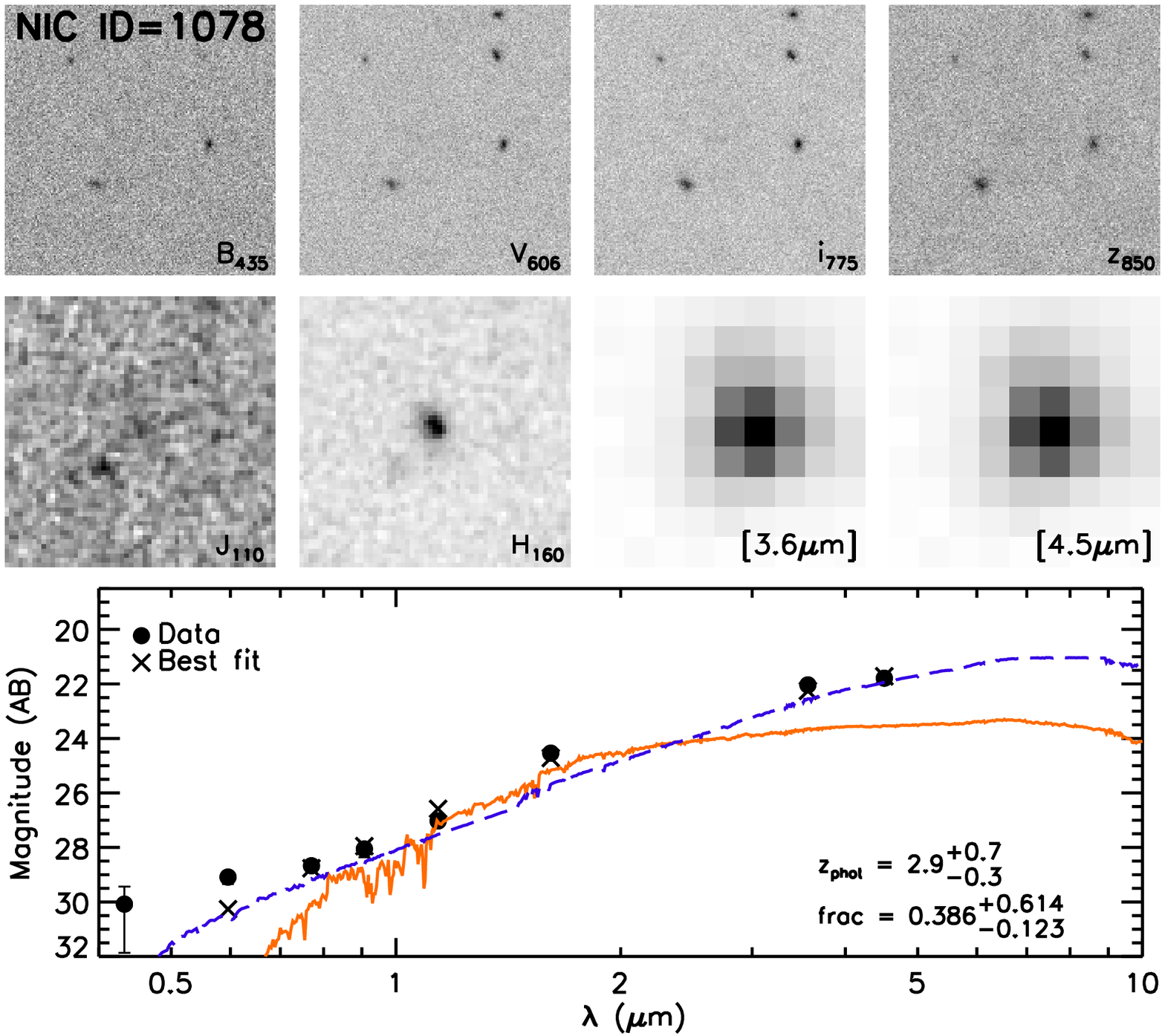}}
      {\includegraphics{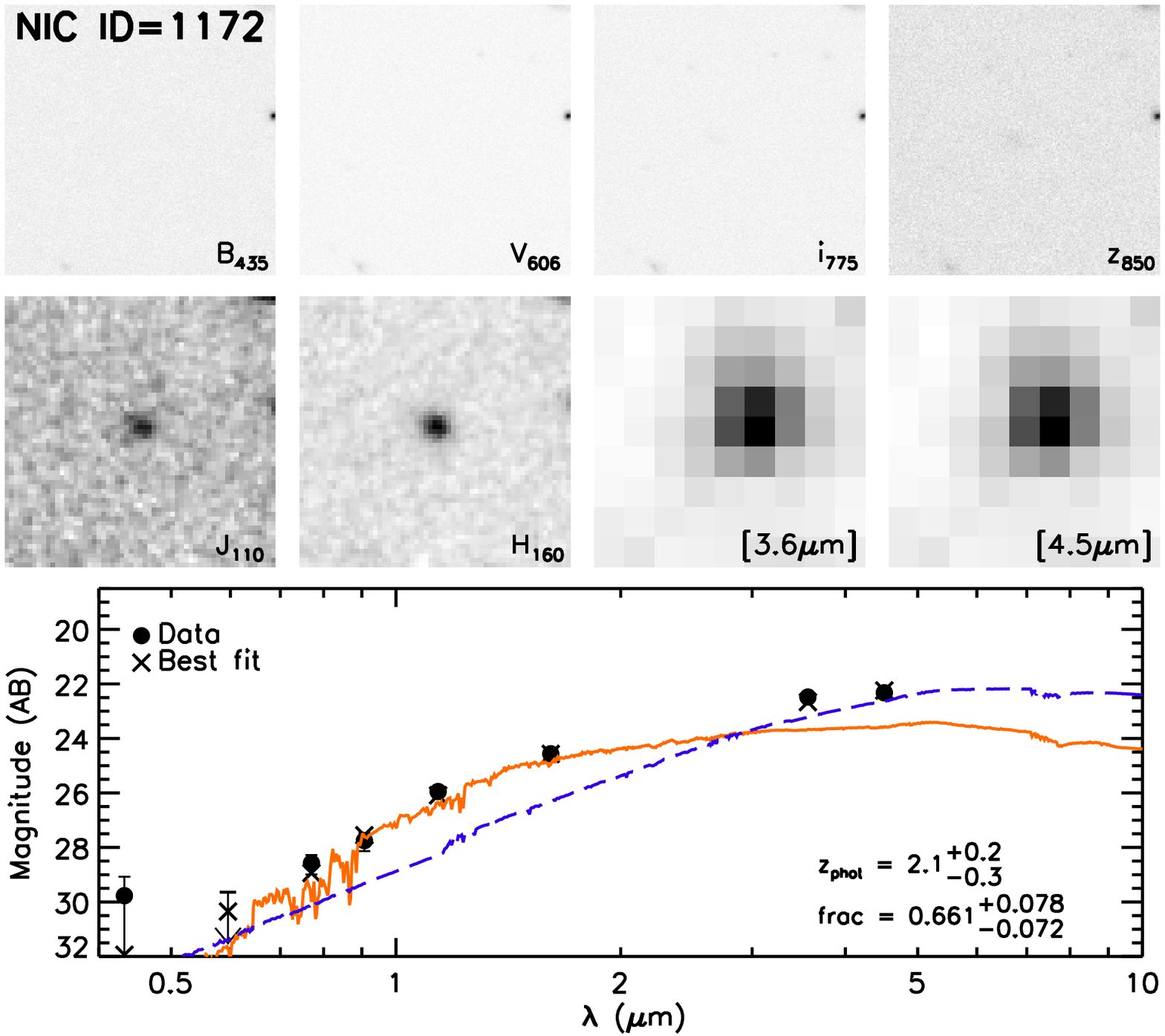}}}
    \scalebox{0.43}{{\includegraphics{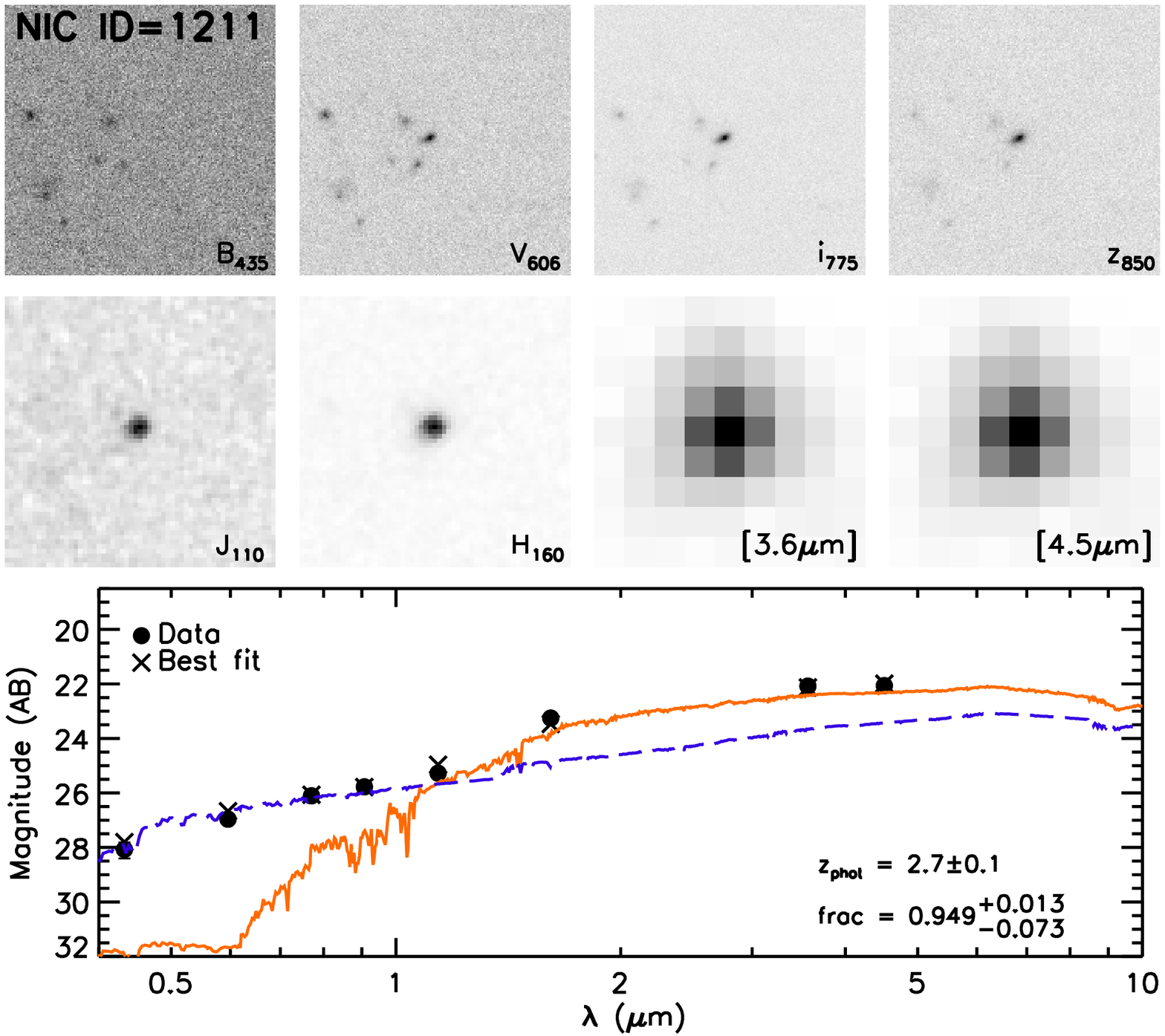}}
      {\includegraphics{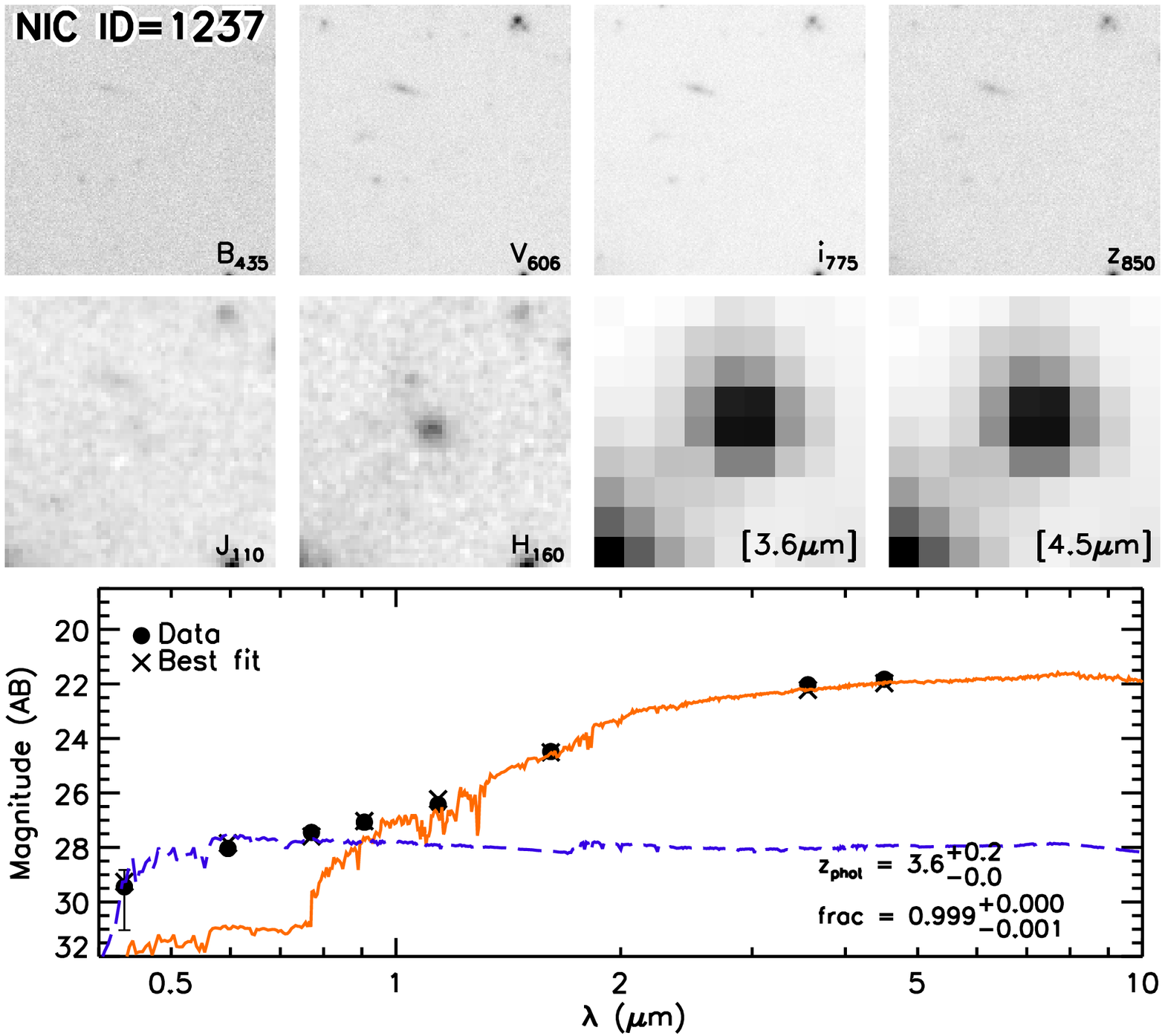}}}
    \caption{Same as figure~\ref{fig2}, arrows indicate $1\sigma$
    upper limits.\label{fig5}}
  \end{center}
\end{figure*}

{\bf \noindent NIC ID $= 921$:} The fitting results are consistent
with an intermediate redshift old stellar population, with
$z_\mathrm{phot} = 1.9 \pm 0.1$ and $f_\mathrm{old} = 0.97$.
\citet{daddi05} measure a consistent result with a spectroscopic
redshift of $z = 1.98$.  \citet{yan04a} also derive a similar
photometric redshift.  Our mass, $8 \times 10^{10}$ \msun\ is in good
agreement with the mass derived by \citet{daddi05}.

{\bf \noindent NIC ID $= 989$:} The resulting model fit is consistent
with a distant galaxy dominated by mass by an old stellar population
and with a small amount of ongoing star formation ($z_\mathrm{phot} =
2.9$ and $f_\mathrm{old} = 0.94$).  Exclusion of the IRAC data yields
a similar model.

{\bf \noindent NIC ID $= 1078$:} The most-likely model parameters
derived from the HST+IRAC data are consistent with a distant,
$z_\mathrm{phot} = 2.9 ^{+0.7}_{-0.3}$, young and dusty stellar
population, with a fitted $f_\mathrm{old} = 0.4$ and $E(B - V) = 1.0$,
although the redshift is not particularly well constrained.  Fitting
the photometry while excluding the IRAC data gives a consistent
redshift, but a very different interpretation of this object.  In this
case the model is dominated by an old stellar population, with
$f_\mathrm{old} = 1.0$.  The relatively bright IRAC flux densities,
taken in conjunction with the HST photometry, are inconsistent with
our models of old stellar populations, which require a flattening of
the SED at wavelengths blueward of about $0.5\micron$ rest-frame.
\citet{chen04} derive a $z_\mathrm{phot} = 2.91$, in good agreement
with our results.  However, they find that this object is consistent
with a E/S$0$ SED, far more evolved than the constant star-forming
model we derive when we include the IRAC data.  \citet{yan04a} select
this object using a $f_{\nu}(3.6\micron)/f_{\nu}(z_{850}) > 20$
selection and derive a $z_\mathrm{phot} = 3.6$, also consistent with
our poorly constrained $z_\mathrm{phot}$.  \citet{yan04a} do not
include this object in their analysis due to the large uncertainties
in their fitted redshift.  However, we note that \citet{yan04a} report
a non-detection for this object (\acsz$> 29.5$~mag; using data from
Yan \& Windhorst 2004c), while we report a detection of \acsz$ =
28.1$~mag.  Upon inspection of the unconvolved ACS z-band image, we
see low surface brightness flux at the location of this object and we
measure a detectable flux in our PSF--convolved image.  Incidentally,
although the analysis of Mobasher et al.\ (2005) supports a very high
redshift interpretation for this object ($z > 6$), subsequent analyses
by Dunlop et al.\ (2007, MNRAS, 376, 1054) and Chary et al.\ (2007)
favor an interpretation that this source resides at substantially
lower redshift $z < 4$, consistent with our interpretation.

{\bf \noindent NIC ID $= 1172$:} This objects' most-likely model
parameters are consistent with an intermediate redshift old stellar
population with a significant amount of UV-bright younger stars:
$z_\mathrm{phot} = 2.1^{+0.2}_{0.3}$ and $f_\mathrm{old} = 0.66$.
\citet{yan04a} derive a $z_\mathrm{phot} = 2.7$, somewhat larger than
our value of $z_\mathrm{phot}$.

{\bf \noindent NIC ID $= 1211$:} The resulting model fit is consistent
with a distant galaxy whose mass is dominated by an old stellar
population.  We derive a $z_\mathrm{phot} = 2.7\pm0.1$ and an
$f_\mathrm{old} = 0.95$.  \citet{chen04}, \citet{daddi05},
\citet{toft05} and \citet{yan04a} derive $z_\mathrm{phot} = 3.43,
2.47, 3.24, 2.8$ respectively.  We suspect the discrepancies between
the different fits result from assumptions about the star-formation
histories of the models and differences in the photometry.  Our
derived mass, $M_{tot} = 1.1 \times 10^{11}$ \msun, is higher than,
but not inconsistent with, the mass derived by \citet{daddi05}.
Although this object is selected by \citet{chen04} they do not report
its mass as they classify it as a dusty starburst.  This object has an
upper limit X-ray flux of $f(0.5-2\mathrm{keV}) = 3.7 \times 10^{-17}$
ergs cm$^{-2}$ s$^{-1}$ \citep{alex03}, implying the possible presence
of an AGN.  However, the rest-frame UV through IR photometry is
consistent with with integrated stellar populations and therefore we
include this object in our sample of distant, evolved galaxies.

{\bf \noindent NIC ID $= 1237$:} This objects' best-fit model is
consistent with a distant, old stellar population: $z_\mathrm{phot} =
3.6$ and $f_\mathrm{old} = 0.999$.  The exclusion if the IRAC data
gives a similar fit, $z_\mathrm{phot} = 3.3$ and $f_\mathrm{old} =
0.998$.  \citet{chen04} derive a much higher redshift for this object,
$z_\mathrm{phot} = 4.25$.  \citet{yan04a} also select this object but
do not consider it due to the fact that their $z_\mathrm{phot} = 3.4$
is very uncertain.  We note that our photometry for this object is
likely biased because of the proximity of a very bright spiral galaxy,
just visible in the lower left corner of fig. 8.  It is likely that
the image convolution causes low-surface-brightness emission for this
objects to enter our aperture.  Furthermore, our photometry could be
biased by the presence of nearby, faint sources in the ACS data that
might be blended with this source at the NICMOS H-band resolution.
This object also has a marginal $24\micron$ detection due to the fact
that it is blended, with $f_{\nu}(24\micron) = 76$~\ujy, which may
call in to question the validity of the best-fit model.  Because of
these reasons we reject this object as a candidate for a distant,
evolved galaxy.

We note that although our photometry is generally in good agreement
with other published values which we are aware of, a few of our
objects have significant differences (notably NIC ID 1078 and 1237).
In the case of NIC ID 1078, we suspect our photometry is appropriate
for our NICMOS-selected catalog.  For NIC 1237, scattered light from a
nearby, large galaxy likely affects our photometry measurements.
Regardless, neither of these objects enters our final sample for
galaxies whose mass is dominated by older stellar populations, so
these do not affect our final conclusions.
 
We derive masses assuming a Salpeter IMF, where the \citet{bc03}
models use upper and lower mass cutoffs of $100$ $\msun$ and $0.1$
$\msun$.  For an SSP \citet{bc03} model, assuming a Chabrier IMF would
decrease our masses by a factor of $1.8$, relative a Salpeter IMF and
for the same assumed upper and lower mass cutoffs listed above.  For a
more detailed discussion of how masses depend on the IMF parameters
such as the slope and the mass cutoffs, see \citet{cp01}.  We note
that \citet{chen04} identify red objects in the HUDF and derive masses
for six early-type candidates, 4 of which are systematically lower
than the masses derived here.  Our overlap objects are $219$, $223$,
$625$, $1078$, and $1237$.  The total H-band magnitudes also differ
for these objects; as expected, there is a rough trend between
increasing flux discrepancy and increasing mass discrepancies.  Object
$223$ has the largest difference in mass, a factor of $\sim 40$.
Objects $219$, $625$ and $1078$ have masses that differ by factors of
18.0, 2.1 and 1.4, respectively, from the \citet{chen04} masses,
although the two estimates for objects 625 and 1078 are entirely
consistent.  Object 1237 is the only for which our mass is lower than
the \citet{chen04} mass, by a factor of $\sim1.2$, but as stated
previously, this likely results from the higher redshift derived by
Chen \& Marzke.

\begin{deluxetable*}{lccccccccc}
\tabletypesize{\scriptsize}
\tablecaption{Comparison between Table 3 and the $\tau$-model fitting}
\tablewidth{0pt}
\tablehead{
\colhead{NIC ID} 
& \colhead{z$_{phot}$} 
& \colhead{z$_{phot}^{\tau}$} 
& \colhead{log M$_{TOT}$/M$_\sun$} 
& \colhead{log M$_{TOT}$/M$_\sun$$^{\tau}$} 
& \colhead{f$_{old}$}
& \colhead{Age (Gyr)$^{\tau}$} 
& \colhead{$\tau$ (Gyr)} 
& \colhead{$\chi_{\nu,best}^{2}$} 
& \colhead{$\chi_{\nu,best}^{2}$$^{\tau}$} 
}
\startdata
30 & 1.8 & 1.5$\pm$0.1 & 11.5 &  11.2$\pm$0.1 & 0.984 & 1.47$\pm$0.36 & 1.19$^{+0.71}_{-0.32}$ & \phn9.2 & \phn8.4  \\
88 & 2.2 & 2.0$^{+0.2}_{-0.5}$ & 10.9 &  10.8$^{+0.1}_{-0.3}$ & 0.718 & 0.08$^{+0.24}_{-0.02}$ & 0.01$^{+0.74}_{-0.01}$ & \phn0.7 & \phn1.3  \\
111 & 1.5 & 0.8$\pm$0.1 & \phn9.6 & \phn7.8$^{+0.0}_{-0.2}$ & 0.914 & 0.14$\pm$0.11 & 0.13$\pm$0.10 & 11.3 & \phn4.5  \\
120 & 3.4 & 2.3$\pm$0.2 & 11.1 &  10.5$^{+0.3}_{-0.4}$ & 0.994 & 0.39$^{+0.92}_{-0.36}$ & 0.91$^{+9.92}_{-0.90}$ & \phn2.0 & \phn1.2  \\
219$^{\bigstar}$ & 2.8 & 2.5$\pm$0.1 & 11.1 &  10.9$\pm$0.1 & 0.982 & 0.80$^{+0.14}_{-0.29}$ & 0.36$\pm$0.16 & \phn2.2 & \phn2.0  \\
223$^{\bigstar}$ & 3.1 & 2.3$^{+0.7}_{-0.1}$ & 11.4 &  11.1$\pm$0.1 & 0.995 & 0.86$^{+0.44}_{-0.19}$ & 0.81$^{+3.09}_{-0.63}$ & \phn0.7 & \phn1.0  \\
269$^{\bigstar}$ & 2.7 & 2.6$\pm$0.1 & 10.4 &  10.1$\pm$0.1 & 0.944 & 0.75$^{+0.19}_{-0.22}$ & 1.68$^{+11.0}_{-1.13}$ & \phn1.9 & \phn2.2  \\
319 & 1.2 & 1.0$^{+0.0}_{-0.1}$ & 10.5 &  10.4$^{+0.1}_{-0.2}$ & 0.344 & 0.08$^{+0.09}_{-0.06}$ & 0.25$^{+3.84}_{-0.25}$ & \phn1.3 & \phn0.9  \\
625$^{\bigstar}$ & 2.7 & 2.7$\pm$0.1 & 10.6 &  10.3$\pm$0.1 & 0.937 & 0.19$^{+0.23}_{-0.08}$ & 0.04$^{+0.06}_{-0.03}$ & \phn3.3 & \phn1.1  \\
706 & 1.3 & 1.2$\pm$0.1 & 10.4 &  10.2$\pm$0.3 & 0.703 & 0.13$^{+0.60}_{-0.09}$ & 0.71$^{+7.17}_{-0.69}$ & \phn4.2 & \phn2.8  \\
921 & 1.9 & 1.9$^{+0.2}_{-1.4}$ & 10.9 &  10.6$^{+0.1}_{-1.8}$ & 0.966 & 0.26$^{+0.22}_{-0.23}$ & 0.01$^{+0.08}_{-0.01}$ & \phn1.6 & \phn1.6  \\
989$^{\bigstar}$ & 2.9 & 2.5$\pm$0.2 & 10.5 &  10.1$^{+0.1}_{-0.3}$ & 0.935 & 0.47$^{+0.34}_{-0.37}$ & 0.68$^{+4.35}_{-0.61}$ & \phn1.3 & \phn2.2  \\
1078 & 2.9 & 2.7$^{+0.2}_{-0.1}$ & 11.0 &  11.2$\pm$0.1 & 0.386 & 0.53$^{+0.75}_{-0.40}$ & 0.12$^{+0.18}_{-0.11}$ & \phn5.8 & \phn1.2  \\
1172 & 2.1 & 1.4$^{+0.3}_{-0.5}$ & 10.6 &  10.4$^{+0.2}_{-0.4}$ & 0.661 & 0.52$^{+0.56}_{-0.41}$ & 0.01$^{+0.11}_{-0.01}$ & \phn1.0 & \phn0.5  \\
1211$^{\bigstar}$ & 2.7 & 2.8$\pm$0.1 & 11.1 &  10.9$\pm$0.1 & 0.949 & 0.34$\pm$0.11 & 0.09$\pm$0.01 & \phn5.0 & \phn1.7  \\
1237 & 3.6 & 2.7$^{+0.2}_{-0.1}$ & 11.2 &  11.2$^{+0.1}_{-0.2}$ & 0.999 & 1.03$^{+0.66}_{-0.41}$ & 0.72$^{+0.54}_{-0.45}$ & \phn1.9 & \phn0.5  \\
\enddata

\tablecomments{The columns with $\tau$-model parameters are indicated
  with a $\tau$ symbol.  The six galaxies with broad-band SEDs
  consistent with distant evolved stellar populations are marked with
  a star symbol.}

\end{deluxetable*}

\subsection{Other models}

To investigate possible systematic dependences due to our modeling, we
explore the effects of changing our model assumptions.

We first consider a single component fit using the same young model
described above, a constant star forming model attenuated with the
\citet{dc00} law.  This model does not characterize our objects well,
yielding very large $\chi^2$ values compared to the two-component base
model, with an average increase in $\chi^2 \sim 55$.

We also test a more complex dust model, the \citet{witt00} SMC shell
geometry dust model.  We try both a single component and a
two-component fit with this dust model.  The single component fit, a
$100$~Myr old constant star forming model with dust as a free
parameter, does not characterize the data well; the best-fit $\chi^2$
values are very large compared to our base model and to the single
component fit using \citet{dc00} dust discussed above.  The two
component fit, the second component being a maximally old SSP model,
results in an average $\chi^2$ difference of $\langle
\chi^2_\mathrm{best,\ base\ model} - \chi^2_\mathrm{best} \rangle =
-1.5$ over all 16 objects, somewhat worse than our base model.  This
dust model gives lower $\chi^2_\mathrm{best}$ values for 9
objects. Object $921$ has the largest $\Delta \chi^2$ of 3.3.  This
dust model assigns systematically higher $E(B - V)$ values, in
particular to objects 30, 625, 921 and 1211, with best-fit $E(B - V)
=$ 1.4, 2.5, 5.3 and 2.8 respectively.  For completeness we note that
the dust grid over which our models span is the grid provided by the
\citet{witt00} look-up tables for this dust law.  It is not the same
grid as that in our base model and it is not a regularly varying grid,
having values of $E(B - V)$ ranging from $0.0$ to $12.3$, in steps of
$\Delta E(B - V) = 0.0875$ at low values and in steps of $\Delta E(B -
V) = 1.75$ at high values.  This difference may hamper our ability to
make a fair comparison with our base model.  However, we do not think
this is the case as none of our original model fits have large values
of $E(B - V)$, all objects having model fit values of $E(B - V) \leq
1.1$.  We note that the best-fit masses differ by less than a factor
of $3.6$ between this model and our base model, excluding object 319,
whose mass increases by about an order of magnitude relative to our
base model.  Similarly, the fractions change by small amounts:
$\langle | \Delta f_\mathrm{old} | \rangle = 0.08$ over the set of 16
objects.  The only objects with a significant change in redshift are
$88$, 319 and 1172 with a $\Delta z_\mathrm{phot} = 0.9$, 1.1 and 1.1
respectively. None of these objects are included in our final sample
of galaxies whose mass is dominated by older stellar populations.

We next consider the possibility that our objects have metallicities
lower than solar.  Again, we perform a similar analysis as outlined
above, with both the young and old models at the same ages as for our
base model and using the \citet{dc00} dust law, but with both
components at a metallicity of $0.2Z_{\odot}$.  For this model, the
average difference in $\chi^2$ is $\langle \chi^2_\mathrm{best,\ base\
model} - \chi^2_\mathrm{best} \rangle = -2.1$, i.e., our base model
does a better job at fitting our object SEDs.  This set of
low-metallicity models yields a lower $\chi^2_\mathrm{best}$ for four
of our objects.  For the entire set of 16 objects the change in
redshift is $\langle | \Delta z_\mathrm{phot} | \rangle= 0.3$, with
object $30$ having by far the largest difference: $z_\mathrm{phot,\
base\ model} - z_\mathrm{phot} = -1.3$.  The best-fit masses change
little, with a variation smaller than a factor of $\sim 3$ for all
objects.  The stellar populations remain consistent for all 16 objects
except 1237, which is fitted as a young starburst with $f_\mathrm{old}
= 0.07$, as opposed to our base model fit with $f_\mathrm{old,\ base\
model} = 0.999$.  This result lends credence to the interpretation
supported by the 24$\micron$ and X-ray data that this object may host
an obscured AGN.  We note that altering the metallicity of the base
model does not significantly change the fitting results for the 6
objects we select as candidates for distant, evolved galaxy candidates
and therefore will not affect our final conclusions.

We also consider the effects of a dusty old population; we fit our
data using the two-component base model outlined above and we use
\citet{dc00} dust to attenuate the old component with $E(B - V) =
0.05$ (or an $A_\mathrm{V} = 0.2$).  For this model we calculate an
average $\chi^2$ change of $\langle \chi^2_\mathrm{best,\ base\ model}
- \chi^2_\mathrm{best} \rangle = 4.1$.  We derive
$\chi^2_\mathrm{best}$ that are lower than those derived with our base
models for $11$ objects (NIC ID = $30$, $88$, $111$, $120$, $269$,
$625$, $706$, $986$, $1078$, $1172$, and $1211$).  For this sub-set,
objects $111$, $625$ $1078$, and $1211$ have the largest differences,
with $\Delta \chi^2 = 32.9$, $9.0$, $11.7$, and $6.2$ respectively.
The other $7$ objects all have $\Delta \chi^2 < 3.3$.  The photometric
redshifts change little, with a $\langle | \Delta z_\mathrm{phot} |
\rangle = 0.15$ for all objects, except $111$, which has a $\Delta
z_\mathrm{phot} = 1.9$.  The mass fractions in old and young stars
change little, with $\langle | \Delta f_\mathrm{old} | \rangle = 0.03$
for all objects except 219 and 1237, both of which have a decrease in
$f_\mathrm{old}$ of $f_\mathrm{old,\ base\ model} - f_\mathrm{old} =
0.9$.  For the 6 objects we consider distant, evolved galaxy
candidates, the derived masses change by less than $50\%$.

Finally, we consider single-component $\tau$-models with varying ages
similar to other studies in the literature.  We summarize the fitting
results here and compare, in more detail, the best-fit $\tau$-model
parameters with our base model results in Table~4.  We generate models
over a grid in ages and $\tau$ values, were $\tau$ is the exponential
timescale for decay in star-formation.  Our age grid values are 0.01,
0.02, 0.04, 0.06, 0.08, 0.10, 0.20, 0.50, 0.70, 1.00, and 2.00 Gyr,
and our $\tau$ grid spans 0.001~Gyr to 20.0~Gyr in quasi-logarithmic
steps.  We use the same redshift and dust grid as that of our base
model and we assume the same dust attenuation law, the \citet{dc00}
law.  We fit the observed SEDs of our selected objects using this
4-dimensional grid of models.  With these models we obtain lower
$\chi^2_\mathrm{best}$ values, relative to our base model, for 11
objects (NIC ID = 30, 111, 120, 219, 319, 625, 706, 1078, 1172, 1211,
and 1237), with objects 111, 1078, and 1211 having the largest
differences of $\Delta \chi^2 = $ 27, 18, and 13, respectively.  For
all 16 objects we measure a change in $\chi^2$ of $\langle
\chi^2_\mathrm{best,\ base\ model} - \chi^2_\mathrm{best} \rangle =
5.0$; if we exclude objects 111, 1078, and 1211, the change in
$\chi^2$ becomes $\langle \chi^2_\mathrm{best,\ base\ model} -
\chi^2_\mathrm{best} \rangle = 1.7$.  The masses of the 11 objects
with lower $\chi^2_\mathrm{best,\tau}$ change by less than a factor 2
with respect to the base model fit, the only exceptions being objects
111 and 120, which have factors of $\sim$60 and $\sim$3 decrease in
mass respectively.  The fitted redshifts do not change significantly,
$\langle z_\mathrm{phot,base\ model} - z_\mathrm{phot} \rangle = 0.3$,
with objects 111, 120, and 1172 having the largest change of $\Delta
z_\mathrm{phot} \sim $ 0.7, 1.0, and 0.7 respectively.  For the entire
sample of 16 it is interesting to note that 14/16 objects have base
model masses that are higher than those derived with the
$\tau$-models.  We consider this to be a consequence of the fact that
our base model is better suited to measure an upper limit on the
masses of our objects; this is in part because the maximally old
component of this model has a limiting mass to light ratio.
Furthermore, we note that for the 6 objects with distant evolved fits,
we derive masses which are between a factor of 1.3 and 2.5 greater
than those derived from the $\tau$-model fits.  Therefore we conclude
that our adopted two-component base model is better suited to place
upper limits on the old stellar population census in our objects than
the $\tau$-model.  We note that the two objects with higher
$\tau$-model masses are 1078 and 1237, which are greater by a factor
of 1.6 and 1.1, respectively, still consistent within the errors with
our base model.

We conclude that a two-component fit is better able to characterize
the data than a one-component fit and that, although some objects are
better fit by other models, in general our base model characterizes
our objects well.  We show that the changes in our fitted redshifts,
total masses and fractions ($f_\mathrm{old}$) are not strongly
dependent on the assumed metallicity and dust properties of our
models.

\section{Discussion and Summary}\label{section:discussion}

We have investigated the physical properties of $32$ galaxies selected
with $(\nicj - \nich) > 1.0$ colors in the HUDF using deep, broadband
imaging from \hst\ and \spitzer.  Exactly half of these objects
(16/32) have unblended IRAC $[3.6\micron]$ and $[4.5\micron]$
photometry.  Here, we focus on the subset of galaxies with IRAC
photometry because the data span a longer wavelength baseline,
allowing us to separate more cleanly the objects into those whose
colors are consistent with dust--enshrouded starbursts and those
dominated by old stars.  Thus, where we derive number densities and
mass densities, we include a factor of 2 correction for incompleteness
due to blending in the IRAC image.  Furthermore, including both the
\spitzer/IRAC data allows us to better constrain the properties of the
galaxies' stellar populations (see further discussion in, e.g.,
Labb\'e et al.\ 2005; Papovich et al.\ 2006).

\subsection{The Properties of Galaxies with Evolved Stellar
Populations at $z\gsim 2.5$}

Based on the full SED fitting, we split the subsample of objects with
IRAC photometry into two broad categories: those objects whose
broad-band SEDs are well represented by our two component model with a
substantial component of evolved stars combined with a young
star-forming component, and those objects whose interpretation changes
substantially by including the IRAC photometry compared to the model
fits to only the ACS and NICMOS data.  For 12/16 (75\%) of the
objects, including the IRAC data give roughly consistent best-fit
models as to the ACS and NICMOS data only (NIC IDs 88, 219, 223, 269,
319, 625, 706, 921, 989, 1172, 1211, and 1237).  Because we are
primarily interested in placing constraints on the number density of
galaxies dominated by passively evolving stellar populations, we focus
here on the subset of galaxies whose model fits have $z \geq 2.5$ and
$f_\mathrm{old} \geq 0.9$.  That is, the models favor a scenario where
the vast majority of the galaxy's stellar mass resides in an older,
passively evolving component.  There are 6 objects satisfying these
conditions (NIC IDs 219, 223, 269, 625, 989, and 1211).

For three of these objects (NIC IDs 223, 625, 989), we have
higher-confidence redshifts due to the presence of a Lyman ``break''
between in the $\acsb -\acsv$ colors.  This occurs for galaxies with
relatively blue rest-frame UV colors, and a characteristic ``break''
in the colors as neutral hydrogen attenuation from the IGM absorbs the
intrinsic flux shortward of 1216~\AA\ (e.g., Madau 1995).  For the ACS
filters, absorption owing to IGM neutral hydrogen affects the
\acsb-band at $z \gsim 3$, and the presence of this break drives the
estimate of the redshifts for these galaxies ($z_\mathrm{phot} \simeq
3.5$, 3.2, and 3.1, respectively).  Interestingly, at $z\sim 3-3.5$
the Balmer/4000~\AA\ is moving through the \nich-band, diminishing the
$J-H$ color. Therefore the fact that these galaxies have $\nicj -
\nich \geq 1$~mag implies they need a relatively large fraction of
their stellar mass in older populations, and in all cases we find best
fits with $f_\mathrm{old} > 0.93$.

Chen \& Marzke (2004) used a NICMOS-selected catalog in the HUDF
combined with photometric redshifts to select six candidates for
massive, red ($\acsi - \nich \geq 2$) early-type galaxies at $z > 3$.
Interestingly, for only 2 of their candidates where we have IRAC data
does our analysis reach the conclusion that old stellar populations
dominate the observed colors (NIC IDs 223, 625).  We find for three of
their candidates (NIC IDs 219, 1078, 1237) that when the IRAC data are
included the SED fits favor solutions where most of the light
originates from dusty, young stellar populations.  For the remaining
candidate of Chen \& Marzke, we do not have IRAC data but our SED fit
to the existing ACS + NICMOS photometry does favor a solution where
the galaxy is dominated by older stellar populations.

For three of our six objects with IRAC photometry, $z\geq 2.7$, and
$f_\mathrm{old} \geq 0.9$ (NIC IDs 269, 989, and 1211), Chen \& Marzke
(2004) do not include these in their sample of evolved galaxies.  We
derive a photometric redshift, $z=2.7$, for NIC ID 269, slightly below
the $z > 2.8$ selection of Chen \& Marzke, which may be the reason
they excluded it.  Chen \& Marzke interpret object NIC ID 1211 to be
dominated by a dusty starburst.  However, the $\nich$ to \mone\ and
\mtwo\ colors imply that while its rest-frame UV light stems from
dust-obscured young stars, most of the mass ($>$88\% at 68\%
confidence) resides in evolved stellar populations (see figure~8;
however, we note that the upper limit on the X-ray detection of this
galaxy may imply that it hosts an AGN, see \S~3.2).  NIC ID 989
appears to have some ongoing star formation (with relatively blue
$\acsb$ to $\acsz$ colors), but older stars appear to dominate the
rest-frame optical and near-IR light (figure~7).  We suspect this
object was excluded from the selection of Chen \& Marzke because of
its relatively blue ACS colors.  Regardless, we find that data at
wavelengths greater than rest-frame 1~\micron\ are required for
interpreting the properties of the stellar populations of red
galaxies.

Parenthetically, we note that there are two objects in the HUDF with
$\nicj - \nich > 2$~mag (NIC IDs 1078, 1211) and one object has $\nicj
- \nich = 1.95$~mag (NIC ID 1237).  All three have IRAC photometry.
These objects have colors comparable to the ``unusual'' red, $\nicj -
\nich$-selected object in the HDF-N reported by Dickinson et al.\
(2000), which has $\nicj - \nich \simeq 2.3$~mag.  Based on our fits
to the ACS, NICMOS, and IRAC photometry of the HUDF objects, we
conclude that two (NIC IDs 1078, 1237) are dominated by dusty
starbursts, while only one (NIC ID 1211) is dominated by a substantial
population of old stars.  Dickinson et al.\ (2000) are unable to
distinguish between these two possibilities (or, for that matter, the
possibility that their object is a $z\approx 12$ Lyman-break-type
galaxy).  This is partly because the HDF-N object is undetected in the
\hst/WFPC2 data, nor did it have longer wavelength data from
\spitzer/IRAC.  Two of the three objects here have \hst/ACS photometry
$\gsim 1$~mag fainter than the detection limit of the WFPC2 data in
the HDF-N.  Therefore, we conclude that one requires \textit{very}
deep \hst\ optical imaging and \spitzer/IRAC imaging to constrain the
nature of these very red, $\nicj - \nich \gsim 2$~mag objects.

None of our six candidates for galaxies with substantial population of
evolved stellar stars is entirely devoid of recent star formation
(i.e., all objects have $f_\mathrm{old} < 1.0$).  Indeed, in the
entire HUDF there are \textit{no} candidates for galaxies consistent
with a purely passively-evolving stellar population formed at much
higher redshift (and in the NICMOS HDF-N, there is only one possible
candidate, see discussion above; Dickinson et al.\ 2000).  Therefore,
such objects are extremely rare if they exist at this redshift.  We
conclude that at $z\gsim 2.5$ even galaxies with a substantial amount
of evolved stellar populations still maintain some ongoing
star-formation.  Papovich et al.\ (2005) argue based on galaxies'
morphologies that at $z\gsim 2$, recurrent star-formation dominates
the rest-frame UV and blue light, which will maintain some level of
homogeneity in the galaxies' internal colors.  Our interpretation here
is consistent with that scenario. However, we note that it appears
that while the young stellar populations dominate the UV and blue
light, in a small fraction of galaxies a substantial population of
older stars dominates at rest-frame wavelengths $\gsim 1$~\micron.

\begin{figure*}[t]
  \begin{center}
    \scalebox{0.48}{{\includegraphics{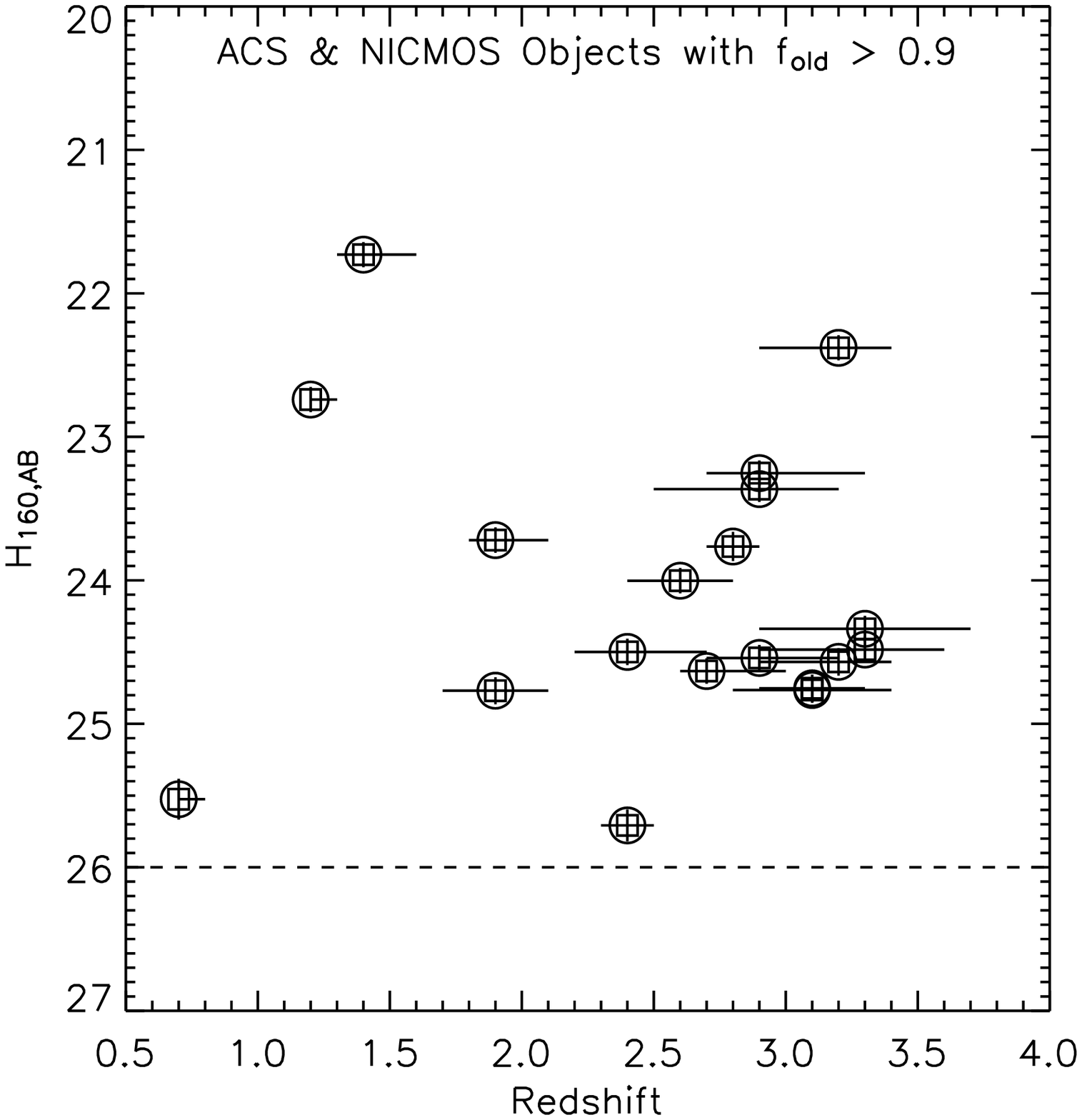}}{\includegraphics{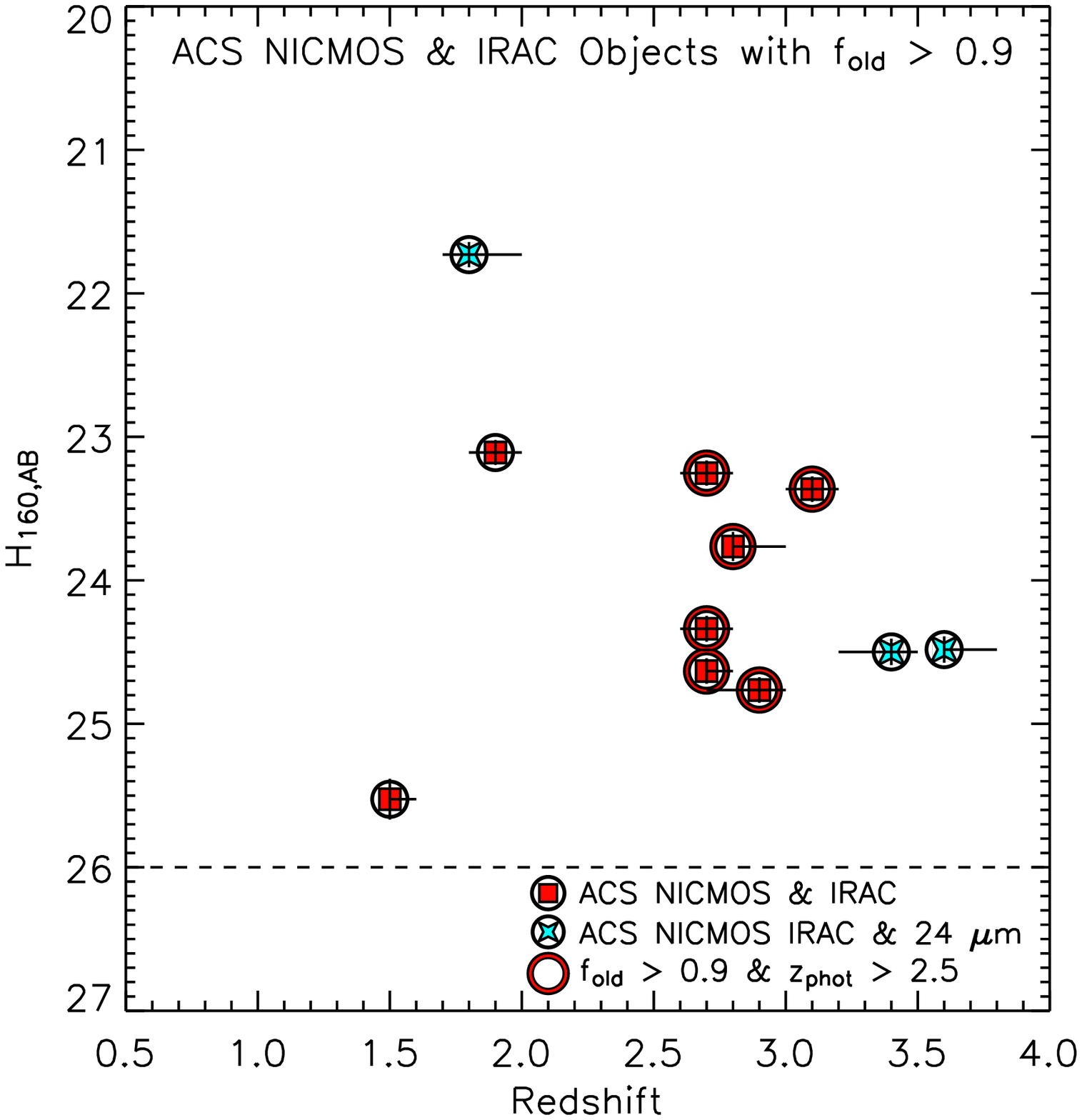}}}
    \caption{$\nich$ magnitudes vs.\ photometric redshifts for the
      subset of $(\nicj-\nich)_\mathrm{AB} > 1.0$~mag objects with
      best-fit $f_\mathrm{old} > 0.9$, for objects without IRAC
      photometry (left panel) and with IRAC photometry (right panel).
      The dashed line shows the $\nich_\mathrm{AB} < 26$ magnitude
      cut.  In the right-hand panel, cyan stars indicate those sources
      among the IRAC--detected subsample with $f_\mathrm{old} > 0.9$
      which have 24~\micron\ detections.  Thick--circles denote
      objects with $f_\mathrm{old} > 0.9$ and $z_\mathrm{phot} > 2.5$}
    \label{fig:co}
  \end{center}
\end{figure*}

\begin{figure*}[t]
  \begin{center}
    \scalebox{0.48}{{\includegraphics{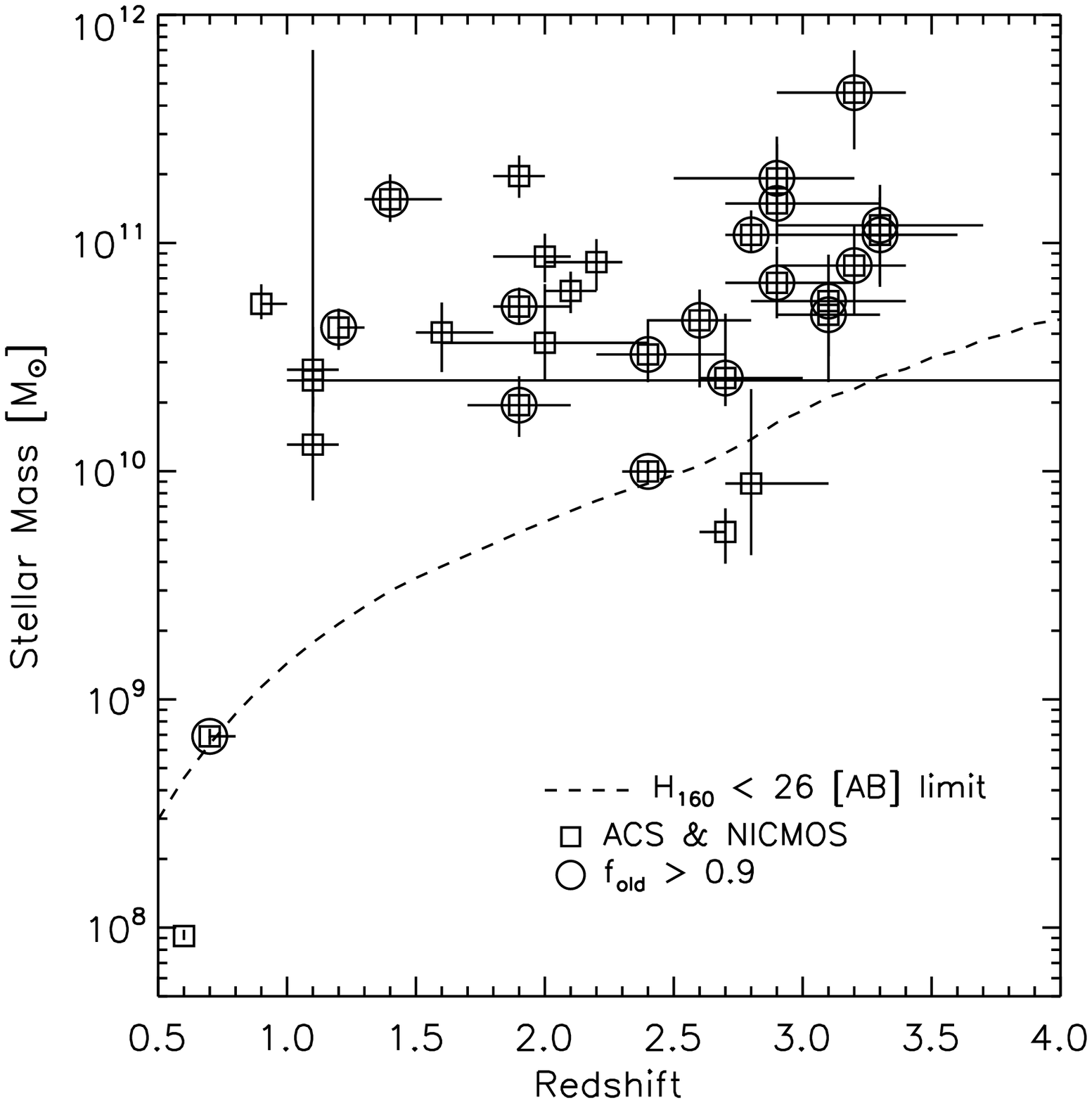}}{\includegraphics{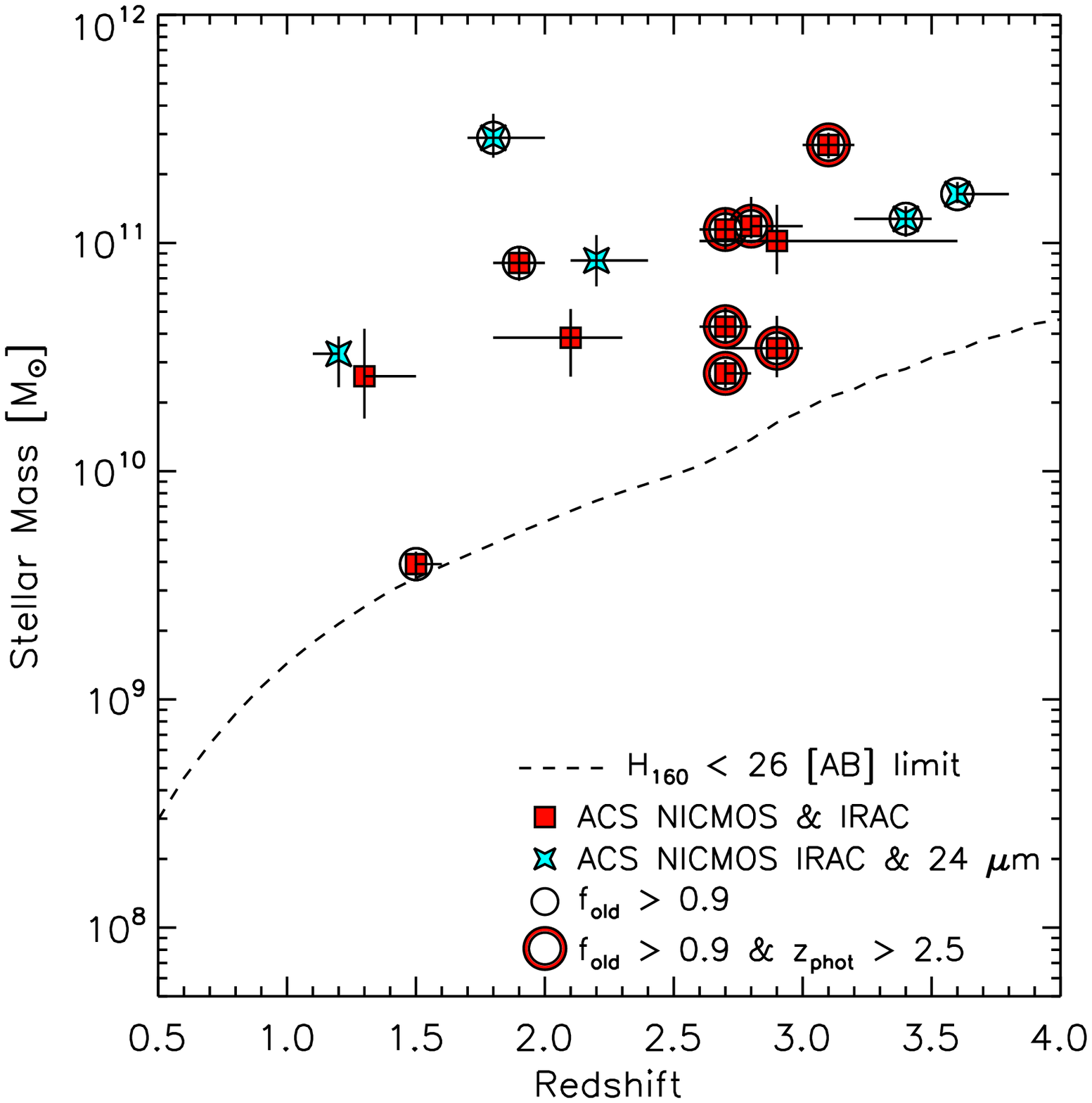}}}
    \caption{Best-fit model masses and photometric redshifts for all
      HUDF objects with $(\nicj-\nich)_\mathrm{AB} > 1.0$~mag (left
      panel) and the subset of these with unblended IRAC photometry
      (right panel).  The dashed line shows the limiting stellar mass
      for galaxies with a stellar population formed in a burst at $z =
      7$ with observed magnitude $\nich = 26.0$.  In both panels the
      open circles indicate objects dominated by old stellar
      populations ($f_\mathrm{old} > 0.9$).  In the right-hand panel,
      cyan stars indicate those sources among the IRAC--detected
      subsample with 24~\micron\ detections.  Thick--circles denote
      those objects with $f_\mathrm{old} > 0.9$ and $z_\mathrm{phot} >
      2.5$, which are the discussion of \S~4.3.}
    \label{fig:co}
  \end{center}
\end{figure*}

\subsection{Galaxies with Evolved Stellar Populations at $z > 1.5$}

With the $\nicj - \nich > 1$~mag selection, we identify massive
galaxies with substantial stellar mass and with SEDs dominated by the
light from passively evolving stellar populations.  Figure~9 shows the
measured $\nich$ magnitudes as a function of redshift for galaxies
with $f_\mathrm{old} > 0.9$, for objects with IRAC photometry (right
panel) and without IRAC photometry (left panel).  Although the data
should be sensitive to such objects down to our selection limit,
$H_{160} < 26$ mag, only one such object has $25 < H_{160} < 26$ mag
--- the rest have $H_{160} < 25$ mag.  Figure~10 shows the derived
stellar masses as a function of galaxy redshift for all galaxies in
the $\nicj - \nich > 1$~mag sample.  The figure shows the stellar
masses derived for the whole sample using only the ACS through NICMOS
data (left panel), and it shows the stellar masses derived including
the IRAC data for the IRAC--detected subsample (right panel).  Objects
with more than 90\% of their stellar mass in evolved stellar
populations are circled in the figure.  There exist objects whose
light (and stellar mass) is dominated by passively evolving stellar
populations with total stellar masses up to and exceeding
$>$$10^{11}$~\msol\ over the entire redshift range.  The existence of
these galaxies is not entirely unexpected, given that such objects
likely dominate the stellar mass density even at $z\sim 2$ (e.g.,
Rudnick et al.\ 2006; van Dokkum et al.\ 2006).

However, there is an apparent dearth of objects whose light is
dominated by passively evolving stellar populations with lower stellar
masses.  For example, none of the 10 IRAC--detected galaxies with
$z_\mathrm{phot} > 1.6$ and with more than 90\% of their stellar
masses evolved stellar populations are fainter than $\nich > 25$~mag
(see figure~9), which is one magnitude above our detection limit.  (We
note there is one object with $z_\mathrm{phot} \simeq 1.5$ and $H =
25.5$~mag, and a derived total stellar mass $4\times 10^{9}$~\msol,
see figure~9.)  Therefore, at these high redshifts, the candidates for
passively evolving galaxies are already massive.  In terms of total
stellar mass, the right-hand panel of figure~10 shows that all the
IRAC--detected objects with $f_\mathrm{old} > 0.9$ and $z > 1.5$ have
stellar masses greater than $\sim$$2 \times 10^{10}$~\msol.  If we
instead consider the results for galaxies from their stellar masses
derived without the IRAC data, there is only a single candidate as a
low-mass galaxy whose mass is dominated by passively evolved stellar
populations (and we consider the stellar masses derived in the old
component without IRAC data to be less robust).

The data should be sensitive to objects dominated by old stellar
populations with stellar masses with $\sim$$3\times 10^{9}$ to
$10^{10}$~\msol (see figure 4 and figure 10).  We note that this is
not a bias related to the sources with IRAC detections.  The IRAC data
from GOODS (5.8 \micron\ flux density of 0.11~$\mu$Jy, $5\sigma$)
would detect a passively evolving stellar population formed at z=7 and
M $> 10^{9.5}$~$\msun$ for all redshifts considered here.

Any plausible mass function for galaxies at $z\sim 2$ would predict a
greater number of galaxies with lower stellar masses than at high
stellar mass.  For example, using the Fontana et al.\ (2006) empirical
galaxy mass function, we would expect roughly double the number of
galaxies with $0.5 - 2.0 \times 10^{10}$~\msol\ compared to those with
$>$$2 \times 10^{10}$~\msol.  This strongly contrasts with the
findings here for the red galaxies dominated by passively evolving
stellar populations.  Therefore, the data suggest that galaxies with
more than 90\% of their stellar mass in evolved stellar populations
have a minimum mass threshold of roughly several times
$10^{10}$~\msol\ at $z\sim 2$.  The lack of such objects with lower
masses may be related to the processes governing stellar mass
build--up in galaxies at these redshifts.

\subsection{Constraints on the Densities of Galaxies with Evolved
Stellar Populations at $2.5 < z < 3.5$}

With the full multiwavelength dataset, we have separated our sample
into those galaxies that are dominated by dust-enshrouded starbursts
and those galaxies whose rest-frame optical and near-IR emission is
consistent with being dominated by a substantial population of evolved
stars (see, e.g., section 4.2, above).  Of the 16 $\nicj - \nich >
1$~mag galaxies with robust IRAC photometry, 6 have best-fit models
favoring a substantial population of evolved stellar populations,
which is 37.5\% of the sample.  Extending this to the full sample of
32 galaxies with $\nicj - \nich > 1$~mag, we estimate there should be
12 such objects in the area of the HUDF.  Using these numbers, we
place constraints on the number and mass density for these types of
objects.  If some of our 6 candidates for evolved stellar populations
turned out to be dusty starbursts, our numbers would be reduced.  Due
to the lower mass to light ratios of younger models, the fitted masses
would also be reduced.  Therefore the densities derived here should be
regarded as upper limits.

The HUDF data reach very deep limits in stellar mass for galaxies
dominated by evolving stellar populations.  For a galaxy whose stellar
mass formed \textit{in situ} at $z_f=7$, the $\nich = 26.0$~mag
detection limit corresponds to a limit in stellar mass ranging from
$6\times 10^9$~\msol\ at $z=2$ to $1\times 10^{11}$~\msol\ at $z=4$
(see figures~4 and 10).  Although the $(\nicj - \nich) > 1$~mag color
selection should be, in principle, sensitive to old stellar
populations at $z \gsim 2$, in practice we find candidates for
galaxies dominated by passively evolving stellar populations
$f_\mathrm{old} > 0.94$ at $z\geq 2.7$.  Likewise, at $z > 3.5$, the
Balmer/4000\AA-breaks move beyond the F160W bandpass, making our color
selection less efficient.  Therefore, to set an upper limit on the
number and mass densities of galaxies dominated by passively evolving
stellar populations, we limit our redshift range to $2.5 < z < 3.5$.
We note that requiring an IRAC detection does not limit our stellar
mass selection; the limiting stellar mass for the IRAC GOODS data
sensitivity limit is always less than that for the NICMOS limit at
these redshifts (see above).

The volume in the area of the HUDF is $1.9 \times 10^{4}$~Mpc$^3$
between $2.5 < z < 3.5$.  Restricting our analysis to only those
galaxies with $2.5 < z < 3.5$ we derive a number density of
3.1$\times$$10^{-4}$~Mpc$^{-3}$, or 6.3$\times$$10^{-4}$~Mpc$^{-3}$
including our 50\% incompleteness.  However, these six objects are not
visible over the whole redshift range.  In the number density
calculation, we compute the effective volume, $V_\mathrm{eff}$ for
each galaxy from $z=2.5$ to $z_\mathrm{eff}$, where $z_\mathrm{eff}$
is the lesser of $z=3.5$ and the maximum redshift to which the galaxy
could be detected.  The number density derived by summing $\Sigma_i
1/V_\mathrm{eff}^{(i)}$ over all $i$ galaxies gives $3.3^{+1.0}_{-1.5}
\times 10^{-4}$~Mpc$^{-3}$, and after accounting for incompleteness,
$n = 6.6^{+2.0}_{-3.0} \times 10^{-4}$~Mpc$^{-3}$.

Assuming these objects will evolve to become present-day, passively
evolving red galaxies, then this number density is significantly lower
than what we would expect using local mass functions.  Using the
$z\sim 0$ mass function parameters listed in table 4 of
\citet{bell03}, we expect the number density of red, early-type
galaxies to be $n(z=0) = 3.5\times 10^{-3}$~Mpc$^{-3}$ in the volume
spanned between $2.5 < z < 3.5$ in the HUDF (and again taking the
redshift-dependent mass limit from
figure~\ref{fig:masslimit})\footnote{Note that \citep{bell03} assumed
a formation redshift, $z_f = 4$, in their mass functions.  This is
only marginally lower than $z_f=7$ used in this analysis.  However,
the change in lookback time from $z\approx 0$ to 4 relative to that
from $z\approx 0$ to 7 corresponds to a change in the mass-to-light
ratio of $M/L_B = 5.8$ to 6.2.  Applying our higher formation redshift
to Bell et al.\ would increase their stellar masses by $\approx 5$\%,
which is negligible compared to other uncertainties.}.  This is larger
than the observed number density (including incompleteness) by roughly
a factor of 5.  Therefore, while we would have expected 65 objects in
the HUDF (i.e., 32 after our $50\%$ completeness) we have found only
6.

This difference is substantially larger than the Poisson noise ($\pm
\sqrt{6}/6$, i.e., 40\%).  It may be that these objects suffer strong
field-to-field clustering (possibly expected for red galaxies, e.g.,
Daddi et al.\ 2003; Quadri et al. 2007).  We estimate the number error
from density variations in the volume between $2.5 < z < 3.5$ in the
HUDF as $0.3\times \sigma_8$.  While we have no measurement for
$\sigma_8$ for these galaxies, based on theoretical models we expect
$\sigma_8 \sim 1.0$ \citep{adelberger05}, implying that the error from
density variations is 30\%.  Summing these errors in quadrature brings
our total error estimate to $\approx$50\% for the expected number of
these galaxies in the HUDF.  Even so, the difference between the 32
expected galaxies and the 6 we find is substantially larger than the
combined errors from counting statistics and cosmic variance.
Therefore, by number, and recalling that these candidates only define
upper limits, the objects in the HUDF which are dominated by evolved
stellar populations make up less than one-third of present-day red,
early-type galaxies.  Thus, at least two-thirds of present-day red,
early-type galaxies must assemble and/or evolve into their current
configuration at $z \lsim 2.5$.

Much of the stellar mass density in galaxies $2.5 < z < 3.5$ could
reside in red-selected galaxies, whose light is dominated by older,
passively evolving stellar populations. For the $\nicj - \nich >
1$~mag galaxies in the HUDF, we derive an upper limit on the stellar
mass density of $\log (\rho^\ast / [M_\odot \mathrm{Mpc}^{-3}] ) =
7.4\pm{0.3}$, including the effect of our redshift-evolving mass
limit, where our errors are derived from a Monte--Carlo simulation of
the data.  When we include our 50\% incompleteness, our total stellar
mass density becomes $\log (\rho^\ast / [M_\odot \mathrm{Mpc}^{-3}] )
= 7.7\pm{0.3}$.  Poisson sampling of the 6 objects gives a factor of
two error and therefore is the dominant source of uncertainty in the
mass density estimate.  We consider this measurement to be an upper
limit on the mass density of distant evolved galaxies for two reasons:
a) it is possible that some of our evolved galaxy candidates are in
fact dusty starbursts, and b) because our two-component model fit
incorporates an a priori upper limit on the mass--to--light ratio in
the form of a maximally old SSP component (see \S 3).

\citet{rudnick03} derive a stellar mass density at $z =
2.80^{+0.40}_{-0.39}$ of log $\rho_* = 7.5^{+0.1}_{-0.1}$ and at $z =
2.01^{+0.40}_{-0.41}$ of log $\rho_* = 7.5^{+0.1}_{-0.1}$.  Because
our mass density provides an upper limit on the mass limit, we
conclude that our derived mass density is consistent those of Rudnick
et al.  In this case, it may be that passively stellar populations
dominate the stellar mass density.  This is consistent with the
analysis of Dickinson et al.\ (2003) who used two--component
star--formation history models to set an upper limit on the stellar
mass density from passively evolving stellar populations hidden
beneath the glare of younger stars in galaxies at $z < 3$.  Dickinson
et al.\ find that at $z\sim 2.5-3$ an upper limit on the stellar mass
density of $\log (\rho^\ast / [\msol \mathrm{Mpc}^{-3}]) =
7.89^{+0.20}_{-0.15}$.  If this reflects reality, then based on our
analysis, almost all of this mass density could reside in faint,
red--selected galaxies at these redshifts.  \citet{vd06} found that at
$z\sim 2$ red-selected galaxies with masses greater than
$10^{11}$~\msol\ dominate the mass density.  Our analysis of the HUDF
galaxies suggests that at lower stellar masses red-selected galaxies
still contribute significantly to the total mass budget.

Lastly, we stress that with only rest-frame UV and optical data, it is
not possible to discriminate between those galaxies dominated by dusty
starbursts and older stellar populations (e.g., Moustakas et al.\
2004; Smail et al.\ 2002).  Although the IRAC data does not resolve
this degeneracy for all galaxies, we show that a sub-set of objects do
have consistent best-fit results and we therefore use these objects in
our analysis.

We wish to thank our colleagues at Steward Observatory for stimulating
conversations.  In particular we would like to thank George Rieke for
useful comments, Xiaohui Fan for generously providing models of the
IGM attenuation, and Karl Gordon for his suggestions and help with the
dust models used in this work.  AMS would also like to thank Richard
Cool, Aleks Diamond-Stanic, Brandon Kelly, Juna Kollmeier, Andy
Marble, John Moustakas, Jane Rigby and Yong Shi for helpful
discussions.  Support for CJP was provided by NASA through the Spitzer
Space Telescope Fellowship Program, through a contract issued by the
Jet Propulsion Laboratory (JPL), California Institute of Technology
under a contract with NASA.  DJE is also supported by an Alfred P.\
Sloan Research Fellowship.

\end{document}